\documentclass[10pt,twocolumn,showpacs,amsmath,amssymb,floatfix,superscriptaddress]{revtex4-2}

\usepackage{graphicx}
\usepackage{dcolumn}
\usepackage{bm}
\usepackage{color}
\usepackage{txfonts}
\usepackage{microtype}
\usepackage{hyperref}
\usepackage[english]{babel}
\usepackage{slashed}
\usepackage{gensymb}
\usepackage{epsfig}
\usepackage[normalem]{ulem}
\usepackage{amsmath}
\usepackage{array}
\usepackage{booktabs}
\allowdisplaybreaks[4]

\begin{document}
\renewcommand{\figurename}{FIG}	

\title{Generation of arbitrarily polarized muon pairs via polarized $e^-e^+$ collision}

\author{Zhi-Wei Lu}
\affiliation{Ministry of Education Key Laboratory for Nonequilibrium Synthesis and Modulation of Condensed Matter, Shaanxi Province Key Laboratory of Quantum Information and Quantum Optoelectronic Devices, School of Physics, Xi'an Jiaotong University, Xi'an 710049, China}
\author{Qian Zhao}\email{zhaoq2019@xjtu.edu.cn}
\affiliation{Ministry of Education Key Laboratory for Nonequilibrium Synthesis and Modulation of Condensed Matter, Shaanxi Province Key Laboratory of Quantum Information and Quantum Optoelectronic Devices, School of Physics, Xi'an Jiaotong University, Xi'an 710049, China}
\author{Feng Wan}
\affiliation{Ministry of Education Key Laboratory for Nonequilibrium Synthesis and Modulation of Condensed Matter, Shaanxi Province Key Laboratory of Quantum Information and Quantum Optoelectronic Devices, School of Physics, Xi'an Jiaotong University, Xi'an 710049, China}
\author{Bo-Chao Liu}
\affiliation{Ministry of Education Key Laboratory for Nonequilibrium Synthesis and Modulation of Condensed Matter, Shaanxi Province Key Laboratory of Quantum Information and Quantum Optoelectronic Devices, School of Physics, Xi'an Jiaotong University, Xi'an 710049, China}
\author{Yong-Sheng Huang}
\affiliation{School of Science, Shenzhen Campus of Sun Yat-sen University, Shenzhen 518107, P.R. China}
\affiliation{Institute of High Energy Physics, Chinese Academy of Sciences, Beijing 100049, China}
\author{Zhong-Feng Xu}
\affiliation{Ministry of Education Key Laboratory for Nonequilibrium Synthesis and Modulation of Condensed Matter, Shaanxi Province Key Laboratory of Quantum Information and Quantum Optoelectronic Devices, School of Physics, Xi'an Jiaotong University, Xi'an 710049, China}
\author{Jian-Xing Li}\email{jianxing@xjtu.edu.cn}
\affiliation{Ministry of Education Key Laboratory for Nonequilibrium Synthesis and Modulation of Condensed Matter, Shaanxi Province Key Laboratory of Quantum Information and Quantum Optoelectronic Devices, School of Physics, Xi'an Jiaotong University, Xi'an 710049, China}

	\date{\today}
	
\begin{abstract}
Generation of arbitrarily spin polarized muon pairs is investigated via polarized $e^-e^+$ collision. We calculate the fully spin-resolved cross section  ${\rm d}\sigma_{e^-e^+\rightarrow \mu^-\mu^+}$  and utilize the Monte Carlo method of binary collision to describe the production and polarization processes of muon pairs. We find that, due to the dependence of  mixed helicities on the scattering angle, arbitrarily polarized muon pairs with both of the longitudinal and transverse spin components can be produced.
The collision of tightly collimated  electron and positron beams with highly longitudinal polarization and nC charge can  generate about $40\%$ muon pairs with longitudinal polarization  and about $60\%$ muon pairs with transverse polarization. The compact high-flux $e^-e^+\rightarrow\mu^-\mu^+$ muon source  could be implemented through the next-generation laser-plasma linear collider, and  would be essential to facilitate the investigation of fundamental physics and the measurement technology in broad areas.
\end{abstract}

	\maketitle
\section{Introduction}\label{one}
Spin polarized muon sources are versatile in the fundamental particle physics, nuclear physics and condensed matter physics \cite{Gorringe2015,Breunlich1989,Yaouanc2011,Blundell1999}. In particle physics, the muon anomalous magnetic moment \cite{Aoyama2020,Borsanyi2021,Abi2021},  rare muon decay \cite{Kuno2001,Gorringe2015,Renga2019}, and  muon-neutrino flavor oscillations  \cite{Alsharo2003,Bandyopadhyay2009,Boscolo2019} are deemed to encode the experimental evidences in search of the new physics beyond the standard model, demanding the unequivocal measurements by using high-flux polarized muon beams. The longitudinally spin polarized (LSP) and transversely spin polarized (TSP)  muon beams can both be used for the  precise measurements of the muon anomalous magnetic moment \cite{Abi2021}, and are crucial for studying the lepton flavor violating \cite{Kuno2001,Blondel2000} and the $CP$ violation \cite{Choi2001,Kittel2008}. In nuclear physics, the high-flux muon beams apply to the muon-catalyzed fusion \cite{Breunlich1989,Toyoda2003,Holmlid2019}, the production of muonic atoms  for probing the nuclear properties \cite{Pachucki2015,Ji2018}, and the muon-capture reaction for producing the nuclear isotope \cite{Hashim2020}. The LSP muons are {used} to study the nuclear structure via the polarized deep-inelastic scattering with above 80\% polarization \cite{Adams1997,Aghasyan2018} or via the muon capture with above 70\% polarization \cite{Brudanin1995,Measday2001}.
Moreover, the highly polarized (above 90\%) slow muons apply to the muon spin relaxation/rotation ($\mu$SR) measurement due to its sensitivity to low spin fluctuation \cite{Yaouanc2011,Blundell1999}, whereby the $\mu$SR effect is significant in quantum materials, radical chemistry, battery materials and elemental analysis \cite{Louca2021}.

The conventional hadronic $\pi^-\pi^+\rightarrow\mu^-\mu^+$ muon source is generated through the proton-nucleon reactions inside a high-$Z$ target, in which the predominantly produced $\pi^\pm$ mesons decay into the polarized muons and, due to  the parity nonconservation, these muons are completely longitudinal polarization \cite{Cook2017,Eaton2017}. The TSP muons can be obtained in the storage ring, in which   the momenta of LSP muons are deflected by 90$^{\circ}$ to obtain the TSP ones \cite{Pavel2004}. The hadronic muon source usually demands a kilometer-scale accelerator to obtain the proton-beam driver with energy above $100$ MeV, and the muons are required to be efficiently captured and rapidly accelerated  from the divergent hadronic showers \cite{Cook2017,Sahai2019}.

At present, with the rapid developments of ultraintense  ultrashort laser techniques, the state-of-the-art laser pulses can achieve
the peak intensities of about $10^{23}$ W/cm$^2$ with a pulse duration of tens of femtoseconds and an energy fluctuation of about 1\% \cite{Danson2019,Yoon2019,Yoon2021}.
Efficient laser-driven plasma accelerators with a gradient exceeding 0.1 TeV/m can provide dense tens-of-MeV proton \cite{Higginson2018, McIlvenny2021} and multi-GeV electron beams \cite{Gonsalves2019} in experiments, and thus have the potential to significantly accelerate the pre-polarized low-energy proton  \cite{Gong2020ene,Jin2020,Li2021pol} and electron beams \cite{Meng2019,Yitong2019,Nie2021}. Moreover,
 the TSP electron (positron) beam can  be directly produced via nonlinear Compton scattering  (nonlinear Breit-Wheeler  pair production)
in a standing-wave   \cite{Sorbo2017,Sorbo2018,Seipt2018}, elliptically polarized \cite{Li2019ult,Wan2020ul}, or bichromatic laser pulses \cite{Seipt2019,Songhh2019,Chen2019po,
Liu2020} due to the quantum radiative spin effects, and the LSP ones can  be produced   via the helicity transfer from circularly polarized $\gamma$ photons in linear \cite{Zhao2021} or nonlinear Breit-Wheeler  processes  \cite{Liyanf2020,Xuekun2021},  which  are generally  pre-produced via Compton scattering \cite{Phuoc2012al, Ligammaray_2020} or bremsstrahlung  \cite{Abbott2016}. Very recently, an all-optical spin rotation method has  been proposed to generate arbitrarily spin polarized (ASP) lepton and ion beams \cite{Wei2022}.
Thus, such a strong laser pulse could be used to drive the direct production of muon pairs through the quantum electrodynamics (QED) processes \cite{Sahai2019}, such as the Bethe-Heitler (BH) and trident muon pair productions via the beam-target interaction:  $\gamma+Z\rightarrow\mu^-+\mu^++Z$ and  $e^-+Z\rightarrow\mu^-+\mu^++e^{-'}+Z$, and the triplet and $e^-e^+$ annihilation muon pair productions via the beam-beam interaction: $e^-+\gamma\rightarrow\mu^-+\mu^++e^-$ and  $e^-+e^+\rightarrow\mu^-+\mu^+$   \cite{Athar2001,Serafini2018,Antonelli2016,
Boscolo2018,Boscolo2020}. In the former, the high-brilliance muon beams can be produced through the interaction of laser-driven GeV  electron beam and high-$Z$ target \cite{2018Rao}. While, in the later, the low-emittance ones can be produced through the collisions of a positron beam of about 45 GeV with target electrons  \cite{Antonelli2016,Boscolo2018,Boscolo2020}.
However, in these proposals, the polarization of  muon pairs is not taken into account. As known, the longitudinal polarization dominates the BH- and trident-like processes  \cite{Motz1969}. According to the helicity configure in the $e^-e^+\rightarrow\mu^-\mu^+$ process \cite{Hikasa1986},  in principle the LSP and TSP muon pairs can both be produced. However, in order to apply the $e^-e^+$ annihilation process as a polarized muon source, one should acquire the comprehensive knowledge of beam effects and polarization characteristics in this leptonic process through the realistic beam-beam collision.

\begin{figure}[!t]	
	\setlength{\abovecaptionskip}{0.2cm}  	
	\centering\includegraphics[width=1\linewidth]{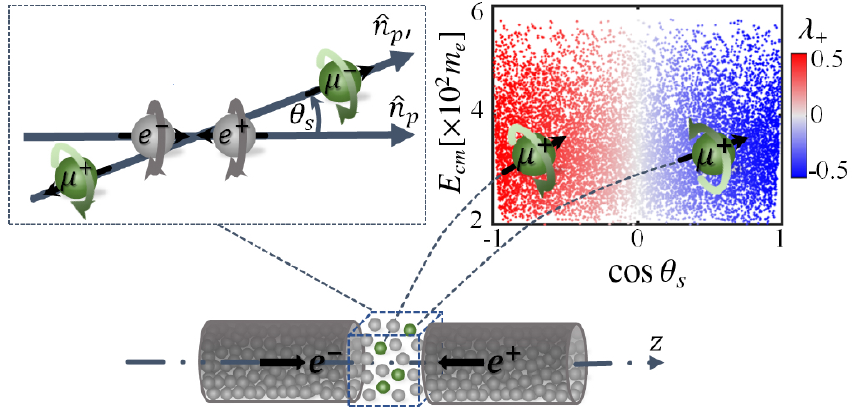}
	\caption{Interaction scenario for  generating ASP muon pairs via polarized $e^-e^+$ collision. Left inset: Schematic helicity configure in the center of mass (c.m.) frame. The black and helical arrows denote the particle momentum direction and  mixed helicity, respectively, $\hat{n}_p$ and $\hat{n}_{p'}$ denote the momentum directions of $e^-$ and $\mu^-$, respectively, and $\theta_s$ is the scattering angle. Right inset:  Helicity distribution of $\mu^{+}$ ($\mu^{-}$ has an opposite distribution) in the plane of c.m. energy $E_{cm}$ and $\cos{\theta_s}$. In our simulation, the electron and positron beams are initialized symmetrically with a transverse Gaussian and longitudinal uniform distribution,  divergence angle $\theta_b=1$ mrad and  uniform energy distribution between $110-360$ MeV in the laboratory frame.}
	\label{fig1}
\end{figure}

In this paper, the generation of ASP muon pairs via the polarized $e^-e^+$ annihilation process \cite{Hikasa1986,Lyuboshitz2009} has been investigated. We calculate the cross section of muon pair production for ASP scattering particles by virtue of the density matrix (see Sec.~\ref{twoA}), and use the fully spin-resolved Monte Carlo (MC) simulation method  of binary collision~\cite{Zhao2021}  to describe the generation and polarization processes of muon pairs (see Sec.~\ref{twoB}). The interaction scenario is illustrated in Fig.~\ref{fig1}.  We find that the collision of polarized electron and positron  with opposite helicity eigenstates annihilates into  polarized muon pair with mixed helicity states, and  the mean helicity distribution of $\mu^+$ ($\mu^-$) is symmetric and  gradually varies with respect to the scattering angle  $\theta_s$ (see Fig.~\ref{fig1} and more details in Figs.~\ref{fig4} and~\ref{fig5}). The longitudinal polarization distributes  near the colliding axis and the transverse polarization distributes approximately perpendicular to the colliding axis. Utilizing laser-driven LSP electron and positron beams, high-flux  ASP muons could be produced directly through ultrahigh-luminosity laser-plasma collider (see Fig.~\ref{fig6} and Table.~\ref{Tab:beam}), which has significant applications in broad areas.

The paper is organized as follows. Sec.~\ref{two} summarizes the methods of theory and numerical simulation. Numerical results and related discussions are given in Sec.~\ref{three}. And, a brief conclusion of this work is presented in Sec.~\ref{four}.

\section{Methods of theory and numerical simulation}\label{two}

The cross section of the $e^-e^+\rightarrow \mu^-\mu^+$ process with definite spin states can be calculated by the helicity scattering amplitude \cite{Hikasa1986,Lyuboshitz2009}. In order to describe the arbitrary polarization in this process, we calculate the fully spin-resolved cross section  by the density matrix (see Sec.~\ref{twoA}), and the differential cross section is shown in Eq.~(\ref{diffCS}).
While, in order to simulate the beam effects of energy and polarization distributions on the muon pair production and build the bridge between theoretical prediction and experimental procedure, we employ the MC simulation strategy of binary collision \cite{Zhao2021} (see Sec. \ref{twoB}), and obtain the transition probabilities for determining the spin states of $\mu^\pm$ [see Eq.~(\ref{prob})].

\subsection{Calculation of the fully spin-resolved cross section of the $e^-e^+\rightarrow \mu^-\mu^+$ process}\label{twoA}

In the calculation of the fully spin-resolved cross section of the muon pair production, we use the Lorentz invariant density matrix to describe the arbitrary polarization (see \cite{Zhao2021}, \S65 and \S87 in \cite{Berestetskii1982}).
Denoting $\bm{\zeta}_\pm$ and $\bm{\zeta}'_\pm$ as mean spin vectors of $e^\pm$ and $\mu^\pm$ (see \S29 in \cite{Berestetskii1982}), respectively, whereby the density matrixes can be formulated by the defined spin 4-vector bases [see Eqs.~(\ref{ele}) and (\ref{mu})].
The obtained cross section in the c.m. frame is
	\begin{eqnarray}\label{Wspin}
	\frac{{\rm d}\sigma}{{\rm d}\Omega}=\frac{r_e^2m_e^4m_{\mu}^2|\bm{p}'|}{16E_{cm}^6|\bm{p}|}|\mathcal{M}|^2~,
	\end{eqnarray}
with $\bm{p}$ and $\bm{p}'$ being the c.m. momenta, $m_e$ and $m_\mu$ being the masses of electron and muon, respectively. $\Omega$ and $r_e$ are the solid angle and classical radius, respectively. The scattering amplitude $|\mathcal{M}|^2$ is expressed as \cite{Berestetskii1982}:
	\begin{eqnarray}\label{amplitude}
	\mid \mathcal{M}\mid^2&=&\left(\bar{v}(p_+,a_+)\gamma^{\alpha}u(p_-,a_-)\bar{u}(p_-,a_-)\gamma^\beta v(p_+,a_+)\right)\nonumber\\
&\times&\left(\bar{u}(p'_-,a'_-)\gamma_{\alpha}v(p'_+,a'_+)\bar{v}(p'_+,a'_+)\gamma_\beta u(p'_-,a'_-)\right) \nonumber\\	
&=&{\rm Tr}^{{\rm Dirac}}(\rho_+\gamma^{\alpha}\rho_-\gamma^{\beta}){\rm Tr}^{{\rm Dirac}}(\rho'_-\gamma_{\alpha}\rho'_+\gamma_{\beta})~,
	\end{eqnarray}
where $\gamma_\alpha$ and $\gamma_\beta$ are the  gamma matrixes, contracted to 4-vectors in `slash' notation below, $p_\pm$ and $p'_\pm$ denote the 4-momenta, and, $a_\pm$ and $a'_\pm$ denote the spin 4-vectors of $e^\pm$ and $\mu^\pm$, respectively. The density matrixes in Eq.~(\ref{amplitude}) are derived from the direct product of the Dirac spinors $u\bar{u}$ and $v\bar{v}$ with the replacements:
\begin{eqnarray}
\left\{
\begin{array}{lr}
u(p_-,a_-)\bar{u}(p_-,a_-)\rightarrow\rho_-~, & \\
v(p_+,a_+)\bar{v}(p_+,a_+)\rightarrow\rho_+~, & \\
u(p'_-,a'_-)\bar{u}(p'_-,a'_-)\rightarrow\rho'_-~, & \\
v(p'_+,a'_+)\bar{v}(p'_+,a'_+)\rightarrow\rho'_+~,&
\end{array}
\right.
\end{eqnarray}
with  $\rho_\pm=1/2(\slashed{p}_\pm\mp m_e)[1-\gamma^5(\slashed{a}_\pm)]$ and $\rho'_\pm=1/2(\slashed{p'}_\pm\mp m_\mu)[1-\gamma^5(\slashed{a'}_\pm)]$. The spin 4-vectors are expanded as $a_\pm=\sum_{j=1}^{3}\zeta_{\pm,j}\hat{e}_j^{\pm}$ and $a'_\pm=\sum_{j=1}^{3}\zeta'_{\pm,j}\hat{u}_j^{\pm}$, where $\hat{e}^{\pm}$ and $\hat{u}^{\pm}$ are the spin 4-vector bases of $e^\pm$ and $\mu^\pm$, respectively. Substituting these expansions into the density matrixes of $e^\pm$ and $\mu^\pm$, one obtains
\begin{eqnarray}\label{rho_e}
\left\{
\begin{array}{lr}
\rho_{\pm,0}=1/2(\slashed{p}_\pm\mp m_e)~,  & \\
\rho_{\pm,j}=-\zeta_{\pm,j}\rho_{\pm,0}\gamma_5\slashed{\hat{e}}_j^\pm, j=1,2,3~,   &
\end{array}
\right.
\end{eqnarray}
\begin{eqnarray}\label{rho_u}
\left\{
\begin{array}{lr}
\rho'_{\pm,0}=1/2(\slashed{p'}_\pm\mp m_\mu)~,  & \\
\rho'_{\pm,j}=-\zeta'_{\pm,j}\rho'_{\pm,0}\gamma_5\slashed{\hat{u}}_j^\pm, j=1,2,3~. &
\end{array}
\right.
\end{eqnarray}
Using the momenta of the scattering particles, a set of orthogonal 4-vectors
\begin{eqnarray}\label{first}
\left\{
\begin{array}{lr}
    Q=p_-+p_+=p'_-+p'_+~, & \\
    M=p'_--p'_+~, & \\
    K_\perp=K-M(p'_-K-p'_+K)/M^2~, & \\
    N=\epsilon^{\lambda}_{\alpha\beta\rho}Q^{\alpha}M^{\beta}K_\perp^{\rho}~, &
\end{array}
\right.
    \end{eqnarray}
is constructed for defining the spin bases of $e^\pm$, and another set of orthogonal 4-vectors
	\begin{eqnarray}\label{second}
	\left\{
	\begin{array}{lr}
	Q=p_-+p_+=p'_-+p'_+~,& \\
	K=p_--p_+~, & \\
	M_\perp=M-K(p'_-K-p'_+K)/K^2~, & \\
	L=\epsilon^{\lambda}_{\alpha\beta\rho}Q^{\alpha}K^{\beta}M_\perp^{\rho}~, &
	\end{array}
	\right.
	\end{eqnarray}
is constructed for defining the spin bases of $\mu^\pm$.  $\epsilon^{\lambda}_{\alpha\beta\rho}$ is the Levi-Civita tensor.
 Thus, it is convenient to define the spin 4-vectors bases for $e^{\pm}$ with
\begin{eqnarray}\label{ele}
\left\{
\begin{array}{lr}
  \hat{e}^{\pm}_0=p_\pm/{m_{e}}~,& \\
  \hat{e}^{\pm}_1=N/{\sqrt{-|N|^2}}~,& \\
  \hat{e}^{\pm}_2=f_1\cdot p_\pm+f_2\cdot p_\mp~,&\\
  \hat{e}^{\pm}_3=f_3\cdot M+f_4\cdot K_\perp~,&
\end{array}
\right.
\end{eqnarray}
and  for $\mu^{\pm}$ with
\begin{eqnarray}\label{mu}
\left\{
\begin{array}{lr}
\hat{u}^{\pm}_0=p'_\pm/m_\mu~,&\\
\hat{u}^{\pm}_1=L/{\sqrt{-|L|^2}}~,&\\
\hat{u}^{\pm}_2=f_5\cdot p'_\pm+f_6\cdot p'_\mp~,&\\
\hat{u}^{\pm}_3=f_7\cdot K+f_8\cdot M_\perp~.
\end{array}
\right.
\end{eqnarray}
 Using the orthogonal relation between the 4-vector spin basis and the normalization condition $|\hat{e}^{\pm}_2|^2=| \hat{e}^{\pm}_3|^2=-1$, the coefficients $f_1,f_2,f_3$ and $f_4$ can be obtained:
\begin{eqnarray}
	\left\{
\begin{array}{lr}
f_1=\frac{\tilde{s}/2-1}{\sqrt{(\tilde{s}/2-1)^2-1}}~, &\\
f_2=-\frac{1}{\sqrt{(\tilde{s}/2-1)^2-1}}~, &\\
f_3=\frac{- \left((\tilde{t}-\tilde{u})^2/(4 m^2-\tilde{s})+\tilde{s}-4\right)}{\sqrt{(\tilde{s}-4) \left((\tilde{s}-4) \left(\tilde{s}-4 m^2\right)-(\tilde{t}-\tilde{u})^2\right)}}~, &\\
f_4=\frac{\tilde{t}-\tilde{u}}{\sqrt{(\tilde{s}-4) \left((\tilde{s}-4) \left(\tilde{s}-4 m^2\right)-(\tilde{t}-\tilde{u})^2\right)}}~, &
\end{array}
\right.
\end{eqnarray}
and the coefficients $f_5, f_6, f_7$ and $f_8$ can be obtained in a similar way as:
\begin{eqnarray}
\left\{
\begin{array}{lr}
f_5=\frac{\tilde{s}/2-m^2}{\sqrt{-(m^4\tilde{s}-\tilde{s}^2m^2/4)}}~,  &\\
f_6=-\frac{m^2}{\sqrt{-(m^4\tilde{s}-\tilde{s}^2m^2/4)}}~, &\\
f_7=\frac{\sqrt{2(2m^2-\tilde{s}/2-(\tilde{t}-\tilde{u})^2/(8-2\tilde{s}))}}{\sqrt{(4-\tilde{s})(\tilde{s}-4m^2)}}~, &\\
f_8=-\frac{(\tilde{t}-\tilde{u})\sqrt{(2m^2-\tilde{s}/2-(\tilde{t}-\tilde{u})^2/(8-2\tilde{s}))}}{(-2m^2+\tilde{s}/2+(\tilde{t}-\tilde{u})^2/(8-2\tilde{s}))\sqrt{2(4-\tilde{s})(\tilde{s}-4m^2)}}~, &
\end{array}
\right.
\end{eqnarray}
where $m=m_\mu/m_e$, and $\tilde{s},\tilde{t}$ and $\tilde{u}$ are normalized Mandelstam invariants defined in Appendix \ref{appB}.

Using the relation d$\Omega$=d(-$\tilde{t}$)d$\psi$/(2$|\bm{p}||\bm{p}'|)$,  substituting Eqs.~(\ref{rho_e}) and (\ref{rho_u}) into Eq.~(\ref{amplitude}) and calculating the Dirac trace, one obtains the differential cross-section of the $e^-e^+\rightarrow \mu^-\mu^+$ process:
\begin{eqnarray}\label{diffCS}
\frac{{\rm d}^2\sigma}{{\rm d}\tilde{t}{\rm d}\psi}&=&\frac{r_e^2}{2\tilde{s}^3(\tilde{s}-4)}(F+\sum_{i=1}^{3}G^-_{i}\zeta'_{-,i}+\sum_{i=1}^{3}G^+_{i}\zeta'_{+,i}\nonumber\\
&+&\sum_{i,j=1}^{3}H_{ij}\zeta'_{-,i}\zeta'_{+,j})~,
\end{eqnarray}
where the functions $F$, $G^{\pm}_{i}$ and $H_{ij}$ are expressed by $\tilde{s},\tilde{t}$ and $\tilde{u}$, and their concrete expressions are presented in Appendix \ref{appB}. After the summation over the final spins, Eq.~(\ref{diffCS}) becomes
\begin{equation}\label{spinsum}
 \frac{{\rm d}^2\bar{\sigma}}{{{\rm d}}\tilde{t}{\rm d}\psi}=\frac{r_e^2}{2\tilde{s}^3(\tilde{s}-4)}F.
 \end{equation}

The spins of the final muon pair $\mu^\pm$ resulted from the scattering process itself are (see \S65 in \cite{Berestetskii1982})
\begin{eqnarray}\label{zeta}
\zeta'^{(f)}_{-,j}=\frac{G^-_{j}}{F},~~~ \zeta'^{(f)}_{+,j}=\frac{G^+_{j}}{F}, j=1,2,3~.
\end{eqnarray}
The spin 3-vectors ${\bm{\zeta}}'^{(f)}_{\pm}$ can be expressed in an arbitrary frame by the definition of  a set of 3-vector basis $\bm{n}'_{\pm}$ \cite{Kotkin1998} as:
\begin{eqnarray}
\bm{\zeta}'^{(f)}_{\pm}&=&\sum_{j=1}^{3} \zeta'^{(f)}_{\pm,j} \bm{n}'_{\pm,j},\\
\bm{n}'_{\pm,j}&=&\hat{\bm{u}}_{j}^{\pm}- \bm{p}'_\pm/(E'_\pm+m_\mu) \hat{u}_{j0}^{\pm},\label{polarization vector}
\end{eqnarray}
with  $\hat{u}_{j0}^{\pm}$ being a time-component of 4-vector $\hat{u}_{j}^{\pm}$ defined in Eq. (\ref{mu}), and, $E'_\pm$ and $\bm{p}'_\pm$ being the energies and momenta of $\mu^{\pm}$ in an arbitrary frame. The 3-vector bases $\bm{n}_{\pm}$ of $e^\pm$ can be defined in the same way.
Thus, the mean helicities of muon pair in an arbitrary frame are expressed as $\lambda_\pm=\bm{\zeta}'^{(f)}_{\pm}\bm{p}'_\pm/(2| \bm{p}'_\pm|)$.

In the c.m. frame, $\zeta'_{\pm,1}$ and $\zeta_{\pm,1}$ are the transverse polarization perpendicular to the scattering plane, $\zeta'_{\pm,3}$ and $\zeta_{\pm,3}$ are the transverse polarization in the scattering plane, and $\zeta'_{\pm,2}$ and $\zeta_{\pm,2}$ are the longitudinal polarization (see details in Sec. \ref{twoB}). Therefore, the helicity amplitudes  can be obtained by setting $\zeta_{\pm,2}=\pm1$ and $\zeta'_{\pm,2}=\pm1$ in Eq.~(\ref{diffCS}) as:
\begin{subequations}\label{4helicity}
\begin{eqnarray}
|\mathcal{M}_{+-\pm\mp}|^2&=&F\pm G^-_{2}\mp G^+_{2}-H_{22}~,\\
|\mathcal{M}_{+-\pm\pm}|^2&=&F\pm G^-_{2}\pm G^+_{2}+H_{22}~.
\end{eqnarray}
\end{subequations}
The corresponding differential cross sections are ${\rm d}\sigma_{+-\pm\mp}$ and ${\rm d}\sigma_{+-\pm\pm}$, where  the subscripts from the first to the fourth in sequence denote the positive (``$|+\rangle$'') or negative (``$|-\rangle$'') helicity eigenstates of  $e^-, e^+, \mu^-$ and $\mu^+$, respectively.

After the integration of Eq. (\ref{spinsum}) over
the azimuth angle $\psi$ and $\tilde{t}$, one obtains the total cross section with
the initial spins $\zeta _{\pm,i}$ of $e^{\pm}$
\begin{eqnarray}\label{intCS}
\bar{\sigma}_{tot}=\frac{r_e^2\pi}{4\tilde{s}^3(\tilde{s}-4)}\tilde{F}~,
\end{eqnarray}
where $\tilde{F}$ is the integration of $F$ with
\begin{eqnarray}\label{F}
\tilde{F}&=&\frac{8\sqrt{(\tilde{s}-4)(\tilde{s}-4m^2)}}{3}(4m^2(\tilde{s}+2)+\tilde{s}(2\tilde{s}+4)\nonumber\\
&+&\zeta_{-,1}\zeta_{+,1}(4m^2(\tilde{s}+2)+\tilde{s}(-\tilde{s}+4))\nonumber\\
&+&\zeta_{-,2}\zeta_{+,2}(4m^2(-\tilde{s}+2)+\tilde{s}(-2\tilde{s}+4))\nonumber\\
&+&\zeta_{-,3}\zeta_{+,3}(4m^2(-\tilde{s}+2)+\tilde{s}(\tilde{s}+4)))~.
\end{eqnarray}

\subsection{MC simulation method}\label{twoB}

The event probabilities are given by $\bar{\sigma}_{tot}$ and the muon pair production is determined by the standard rejection method (see Appendix \ref{appA}). The single collision  is completed by a paired $e^-$ and $e^+$ using the no-time-counter method in a 3-dimensional cell in the laboratory frame \cite{Gaudio2020}. The momenta of the paired $e^{\pm}$ are transformed into the c.m. frame by the Lorentz boost along the direction of the c.m. velocity $\bm{\beta}_{cm}=(\bm{p}_-+ \bm{p}_+)/(E_-+E_+)$, where $\bm{p}_\pm$ and $E_\pm$ are the momenta and energies of $e^-$ and $e^+$ in the laboratory frame, respectively. The obtained momentum $\bm{p}$ and energy $E_{cm}$ in the c.m. frame are substituted into the differential cross section to determine the c.m. momentum $\bm{p}'$ of muon, then the energies $E'_\pm$ and momenta $\bm{p}'_{\pm}$ of $\mu^{\pm}$ in the laboratory frame are obtained by the inverse Lorentz boost.

\begin{figure}[!t]
	\setlength{\abovecaptionskip}{-0.2cm}
	\centering\includegraphics[width=0.8\linewidth]{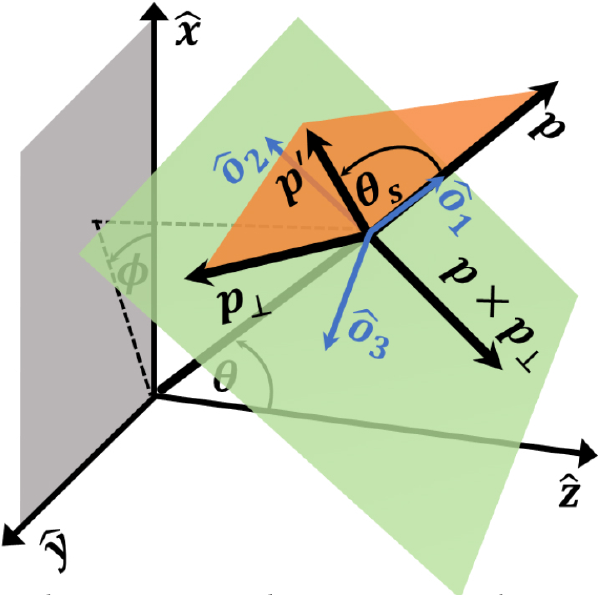}
	\caption{Coordinate system of the muon pair production in the c.m. frame. $\theta$ and $\phi$ are the polar and azimuth angles of the electron momentum $\bm{p}$, respectively, and ($\hat{\bm{o}}_1,\hat{\bm{o}}_2, \hat{\bm{o}}_3$) are the spherical coordinates. $\bm{p}_{\perp}$ in the $\hat{\bm{o}}_2-\hat{\bm{o}}_3$ plane is perpendicular to $\bm{p}$, and $\bm{p}'$ in the  $\bm{p}-\bm{p}_{\perp}$ plane is the momentum of $\mu^-$. }
	\label{fig2}
\end{figure}

Since the reaction is described in the c.m. frame, the coordinate system of the scattering particles in the c.m. frame should be constructed and is sketched in Fig. \ref{fig2}. $\bm{p}_{\perp}$ is determined by generating a random azimuth angle in the $\hat{\bm{o}}_2-\hat{\bm{o}}_3$ plane. The direction of $\bm{p}'$ is determined by the scattering angle $\theta_s$  (see the details in Appendix \ref{appA}), and the later is obtained by solving $\bar{\sigma}_{\theta}/\bar{\sigma}_{tot}=|R_1|$, where $R_1$ is a uniform random number between -1 and 1, and
\begin{eqnarray}\label{partial}
\bar{\sigma}_{\theta}=\sqrt{(\tilde{s}-4)(\tilde{s}-4m^2)}/2\int_{-|\rm cos{\theta_s}|}^{|\rm cos{\theta_s}|} {\rm d}\bar{\sigma}~,
\end{eqnarray}
 which results in
\begin{eqnarray}\label{sigtheta}
\bar{\sigma}_{\theta}&=&\sqrt{(\tilde{s}-4)(\tilde{s}-4m^2)} \left\{4\tilde{s}|\cos{\theta_s}|\left[4m^2(\zeta _{-,1} \zeta _{+,1}-\zeta _{-,2} \zeta _{+,2}\right.\right.\nonumber\\
&-&\zeta _{-,3} \zeta _{+,3}+1)+\tilde{s}(-\zeta _{-,1} \zeta _{+,1}-\zeta _{-,2} \zeta _{+,2}+\zeta _{-,3} \zeta _{+,3}+1) \nonumber\\
&+&\left.4(\zeta _{-,1} \zeta _{+,1}+1+\zeta _{-,2} \zeta _{+,2}+\zeta _{-,3} \zeta _{+,3})\right]\nonumber\\
&-&\frac{4}{3}|\cos{\theta_s}|^3(4m^2-\tilde{s})\left[\tilde{s}(\zeta _{-,1} \zeta _{+,1}-\zeta _{-,2} \zeta _{+,2}-\zeta _{-,3} \zeta _{+,3}+1)\right.\nonumber\\
&-&\left.\left.4(\zeta _{-,1} \zeta _{+,1}+\zeta _{-,2} \zeta _{+,2}+\zeta _{-,3} \zeta _{+,3}+1)\right]\right\} ~.
\end{eqnarray}

In the c.m. frame, the spin 3-vector basis of $e^-$ ($e^+$) $(\bm{n}_{-,1},\bm{n}_{-,2},\bm{n}_{-,3})$
corresponds to the directions of $(\bm{p}_{\perp}\times\bm{p}$, $\bm{p}$, $-\bm{p}_{\perp})$, respectively, and
the defined 3-vector bases for $\mu^\pm$ in Eq.~(\ref{polarization vector}) are simplified as the spherical coordinates defined by $\bm{p}'$ (see Fig.\ref{fig2}). Defining $\zeta'_{\parallel}$ ($\zeta_{\parallel}$) and $\zeta'_{\perp}$ ($\zeta_{\perp}$) as the longitudinal and transverse polarization of muon (electron), respectively, which are expressed as $\zeta'_{\parallel}=\zeta'^{(f)}_{-,2}$ and  $\zeta'_{\perp}=\sqrt{\left(\zeta'^{(f)}_{-,1}\right)^2+\left(\zeta'^{(f)}_{-,3}\right)^2}$ , and the total polarization is $|\bm{\zeta}'^{(f)}_{-}|$.

The projection of the $\mu^\pm$ spin axises $\bm{\zeta}'_\pm$ onto the defined spin states $\pm\bm{\zeta}_\pm^{(d)}$ (unit vector) of a detector is simulated by the MC method, which consists of four transition probabilities
\begin{subequations}\label{prob}
	\begin{align}
	W^{\uparrow\uparrow}&=&\int {{\rm d}}\Omega(0.25+\bm{\zeta}'^{(f)}_+\bm{\zeta}_+^{(d)}+\bm{\zeta}'^{(f)}_-\bm{\zeta}_-^{(d)}+(\bm{\zeta}_-^{(d)})^TH\bm{\zeta}_+^{(d)}),\\
	W^{\uparrow\downarrow}&=&\int {{\rm d}}\Omega(0.25+\bm{\zeta}'^{(f)}_+\bm{\zeta}_+^{(d)}-\bm{\zeta}'^{(f)}_-\bm{\zeta}_-^{(d)}-(\bm{\zeta}_-^{(d)})^TH\bm{\zeta}_+^{(d)}),\\
	W^{\downarrow\uparrow}&=&\int {{\rm d}}\Omega(0.25-\bm{\zeta}'^{(f)}_+\bm{\zeta}_+^{(d)}+\bm{\zeta}'^{(f)}_-\bm{\zeta}_-^{(d)}-(\bm{\zeta}_-^{(d)})^TH\bm{\zeta}_+^{(d)}),\\
	W^{\downarrow\downarrow}&=&\int {{\rm d}}\Omega(0.25-\bm{\zeta}'^{(f)}_+\bm{\zeta}_+^{(d)}-\bm{\zeta}'^{(f)}_-\bm{\zeta}_-^{(d)}+(\bm{\zeta}_-^{(d)})^TH\bm{\zeta}_+^{(d)}).
	\end{align}
\end{subequations}
Each transition probability determines the definite sign of the observed spin components $\zeta_{\pm,i}^{(d)}=\zeta'^{(f)}_{\pm,i}/|\bm{\zeta}'^{(f)}_{\pm}|$  ($i=1,2,3$), and thus determines the parallel or anti-parallel projection of the spin axis. Defining $\bm{\zeta}_\pm^{(d)}$ along the directions of $\bm{\zeta}'_\pm$ leads to the total polarization of  produced $\mu^\pm$ beams
\begin{align}\label{polbeam}
P^{(\mu)}_{tot}=\sqrt{(\bar{\zeta}_{\pm,1}^{(d)})^2+(\bar{\zeta}_{\pm,2}^{(d)})^2+(\bar{\zeta}_{\pm,3}^{(d)})^2},
\end{align}
with the averaged components $\bar{\zeta}_{\pm,i}^{(d)}$ over the particle number. If one defines $\bm{\zeta}_\pm^{(d)}$ as the directions or perpendicular directions of the $\mu^\pm$ momenta, the statistical beam polarization in Eq.~(\ref{polbeam}) leads to the longitudinal polarization ($P^{(\mu)}_{\parallel}$) or transverse polarization ($P^{(\mu)}_{\perp}$). The numerical simulation method is described in detail in Appendix \ref{appA}.

In the simulations, we consider that the
 energy distribution in the given electron and positron beams is spatially homogeneous and  defined by  a characteristic constant, then the muon yield $N_{sc}$ in a single $e^-e^+$ collision can be written as \cite{Esnault2021}
 \begin{equation}
   N_{sc}=\iiiint\frac{{{\rm d}}n_\mu}{{{\rm d}}\tau}{{\rm d}^3}V {{\rm d}}\tau=\mathcal{L}_{e^-e^+}\times\bar\sigma_{tot}^{int},
 \end{equation}
 in term of the single collision geometric luminosity  $\mathcal{L}_{e^-e^+}$ of the beam-beam collision \cite{Herr2006}, where ${{\rm d}}^3V$ and ${{\rm d}}\tau$ are the infinitesimal volume and time, respectively, and an integrated cross section $\bar\sigma_{tot}^{int}$ obtained by coupling the energy distribution functions of electron and positron beams to the total cross section $\bar\sigma_{tot}$. In our simulation, instead of $\mathcal{L}_{e^-e^+}$ which measures the ability of an collider producing the required number of interactions,   the $e^-e^+$ collisions  are sampled by the Thomson cross section $\sigma_T$, which leads to the maximum number of collision $N_{max}$ (see Appendix \ref{appA}). Thus,  the muon yield can also be estimated as
 \begin{equation}
   N_{sc}=\frac{N_{max}}{\sigma_T}\times\bar{\sigma}_{tot}^{int}.
 \end{equation}
In the simulation, the energy-coupled $\bar{\sigma}_{tot}^{int}$  is considered into the MC method, and for the electron and positron beams with specific parameters, the total muon yield is thus estimated as $N_{tot}=f_{rep}N_{sc}$, where $f_{rep}$ is the repetition frequency of a collider.
Since $N_{max}$ is determined by the beam density, the quantity $f_{rep}N_{max}$ in the expression of $N_{tot}$ can be interpreted as the equivalent maximum number of collision resulted from a single beam-beam collision with an assumed charge.

\section{Simulation results and discussions}\label{three}

\begin{figure}[!t]
	\setlength{\abovecaptionskip}{-0.2cm}
	\centering\includegraphics[width=1\linewidth]{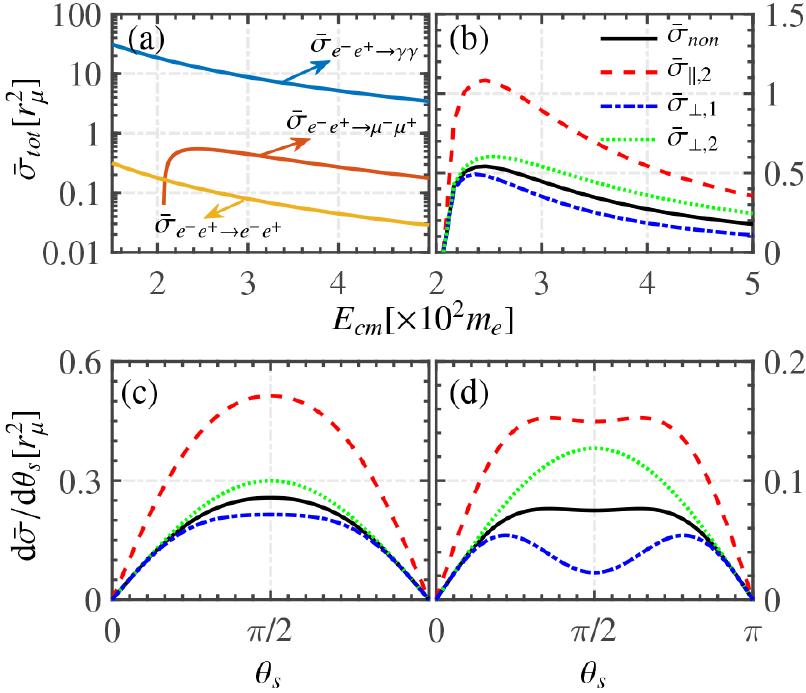}
	\caption{(a) The total cross sections of three processes in the $e^-e^+$ interaction. (b) $\bar{\sigma}_{tot}$ vs $E_{cm}$ calculated from Eq.~(\ref{intCS}) with four initial spin states: nonpolarized $\bar{\sigma}_{non}$, LSP $\bar{\sigma}_{\parallel,2}$ with $\zeta_{-,2}\zeta_{+,2}=-1$, and TSP $\bar{\sigma}_{\perp,1}$ and $\bar{\sigma}_{\perp,2}$ with $\zeta_{-,1}\zeta_{+,1}=1$ and $\zeta_{-,1}\zeta_{+,1}=-1$, respectively. (c) and (d): The differential cross sections vs $\theta_s$ at the peak energies $E_{p}$ and $2E_{p}$, respectively. The line types in (c) and (d) have the same meanings as those in (b) but for ${\rm d}\bar{\sigma}_{non}/{\rm d}\theta_s$, ${\rm d}\bar{\sigma}_{\parallel,2}/{\rm d}\theta_s$, ${\rm d}\bar{\sigma}_{\perp,1}/{\rm d}\theta_s$ and ${\rm d}\bar{\sigma}_{\perp,2}/{\rm d}\theta_s$ calculated from Eq.~(\ref{spinsum}). The unit of the cross section is $r_\mu^2$ with $r_\mu=r_e/m$. }
	\label{fig3}
\end{figure}

\begin{figure}[!t]
	\setlength{\abovecaptionskip}{-0.2cm}
	\centering\includegraphics[width=1\linewidth]{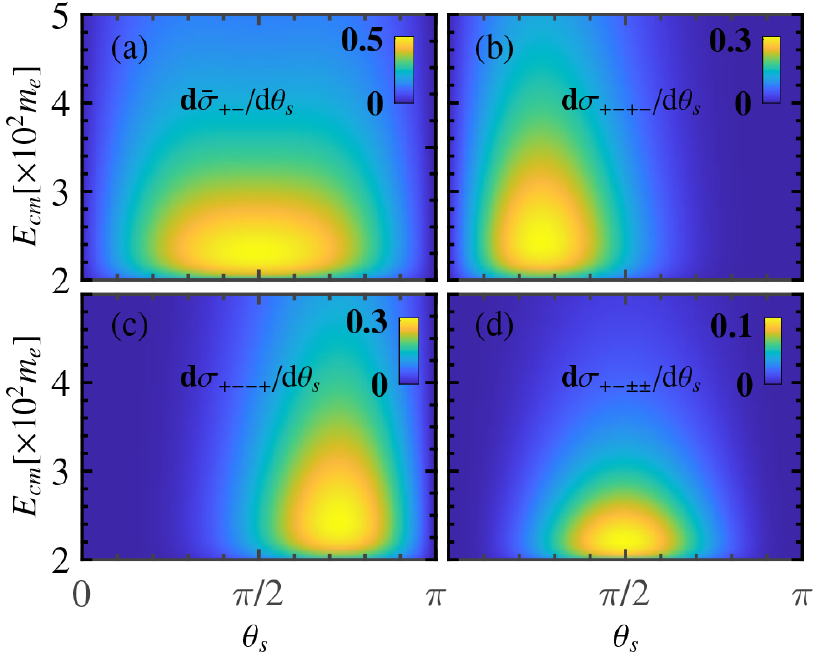}
	\caption{Distributions of differential cross section  in the plane of  $\theta_s$ and  $E_{cm}$: (a) with summarized final spins  ${\rm d}\bar{\sigma}_{+-}/{\rm d}\theta_s$ [calculated via Eq.~(\ref{spinsum})]; (b)-(d) with different helicity channels [calculated via Eq.~(\ref{4helicity})].}
	\label{fig4}
\end{figure}

\begin{figure}[b]
	\setlength{\abovecaptionskip}{-0.2cm}
	\centering\includegraphics[width=1\linewidth]{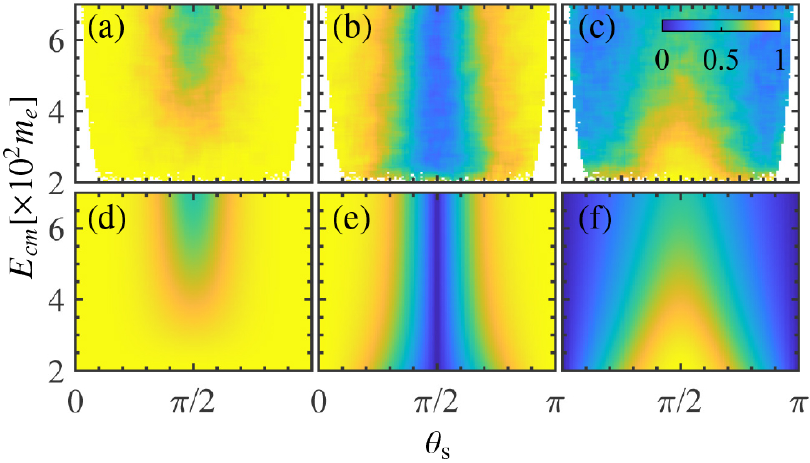}
	\caption {(a)-(c) Polarization distributions  produced by colliding electron and positron beams with opposite helicity eigenstates,  corresponding to $P^{(\mu)}_{tot}$, $P^{(\mu)}_{\parallel}$ and $P^{(\mu)}_{\perp}$, respectively, calculated via the MC method. (d)-(f): Similar to (a)-(c), respectively, but via the analytical calculation in Eq.~(\ref{zeta}). The simulation parameters are the same as those in Fig.~\ref{fig1}.}
	\label{fig5}
\end{figure}

\begin{figure}[!t]
	\setlength{\abovecaptionskip}{-0.2cm}
	\centering\includegraphics[width=1\linewidth]{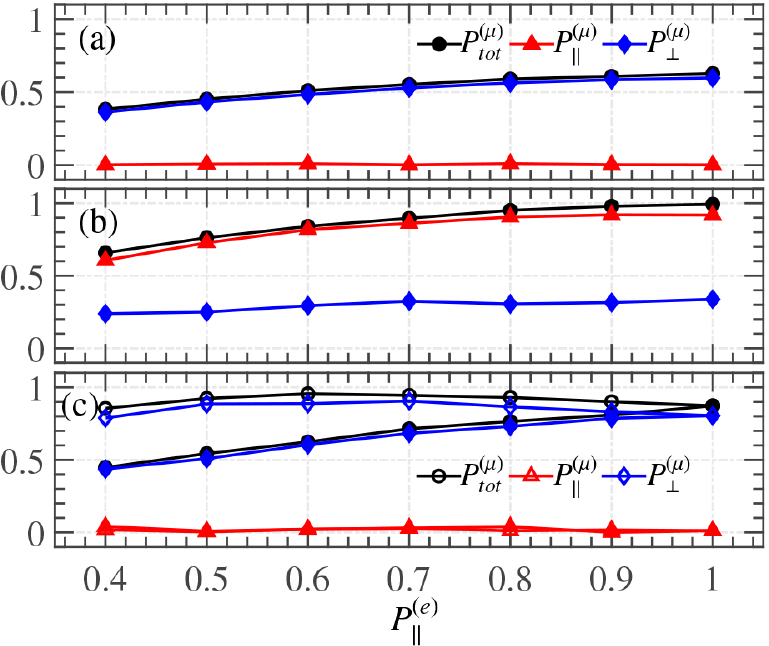}
	\caption{(a)-(c): Polarization of muons  vs  $P^{(e)}_{\parallel}$  for all the muons, the  muons beamed into $0<\theta<0.2\pi$ and $0.3\pi<\theta<0.7\pi$, respectively. The filled marks denote the results produced from the collisions with anti-parallel transverse polarization $\bm{\zeta}_{+,\perp}=-\bm{\zeta}_{-,\perp}$, and the hollow marks in (c) denote the results produced from the collisions with parallel transverse polarization $\bm{\zeta}_{+,\perp}=\bm{\zeta}_{-,\perp}$. Other  parameters are the same as those in Fig.~\ref{fig1} but having an exponential energy distribution with an average energy $E_{av}=125$ MeV in the laboratory frame.}
	\label{fig6}
\end{figure}

In this section, we first present the impact of initial spins of $e^\pm$ on the cross section with summarized final spins (see Fig.~\ref{fig3}), then, the polarization mechanism in the muon pair production is analyzed via the helicity amplitudes (see Figs.~\ref{fig4} and \ref{fig5}). Finally, by varying the initial polarization of  electron and positron beams, the polarization properties of the muon beams are obtained (see Fig.~\ref{fig6}), and the feasible beam parameters for producing the required polarized muon pairs are examined (see Table.~\ref{Tab:beam}).

As known, the annihilation process $e^-e^+\rightarrow\gamma\gamma$ and elastic scattering process $e^-e^+\rightarrow e^-e^+$ also occur in the $e^-e^+$ collision, and the corresponding total cross sections are shown in Fig.~\ref{fig3}(a). These three processes are competitive in a single collision, and the annihilation process is the most probable reaction. In the simulation, these three processes are considered into the rejection procedure simultaneously, the $\gamma$-photon yield produced in the annihilation process is more than ten times larger than the muon pair yield, and this difference is consistent with that between the cross sections (see the comparison in Appendix A). The influence of the elastic scattering process on the initial polarization of electron and positron beams is negligible due to the smallest cross section $\bar{\sigma}_{e^-e^+\rightarrow e^-e^+}$.

The effects of initial spin states of $e^\pm$ on the spin-summarized cross section of the muon pair production are shown in Figs.~\ref{fig3}(b)-(d). Here one has $\bar{\sigma}_{ non}=(\bar{\sigma}_{\parallel,1}+\bar{\sigma}_{\parallel,2})/2=(\bar{\sigma}_{\perp,1}+\bar{\sigma}_{\perp,2})/2$. Note that the component $\bar{\sigma}_{\parallel,1}$ with $\zeta_{-,2}\zeta_{+,2}=1$ equals zero because the corresponding helicity channel leads to the nonexistent virtual photon state and is forbidden. Therefore, $\bar{\sigma}_{\parallel,2}$ is twice  $\bar{\sigma}_{non}$ [see Fig.~\ref{fig3}(b)]. For the spin-summarized differential cross section, its distribution with respect to $\theta_s$ presents the different energy-dependence for different spin states of $e^\pm$. As $E_{cm}$ increases,  ${\rm d}\bar{\sigma}_{non}/{\rm d}\theta_s$ and ${\rm d}\bar{\sigma}_{\parallel,2}/{\rm d}\theta_s$ vary from the cosine angular spectra to the flat-top angular spectra, and the cosine-shaped ${\rm d}\bar{\sigma}_{\perp,2}/{\rm d}\theta_s$ increases against ${\rm d}\bar{\sigma}_{\perp,1}/{\rm d}\theta_s$ decreasing with two peaks [see Figs.~\ref{fig3}(c) and (d)]. $\bar{\sigma}_{tot}$ and ${\rm d}\bar{\sigma}/{\rm d}\theta_s$ indicate the optimal energy distribution of the colliding electron and positron beams for a high muon yield, and the featured angle spectra corresponding to different energy and initial polarization, respectively.

Considering the spins of the muon pair in Eq.~(\ref{diffCS}) with arbitrary $\bm{\zeta}'_+$ and $\bm{\zeta}'_-$, the polarization of  muons with mixed spin states can be clarified through the helicity distribution.
The mean helicities of muon pair originate from the superposition of various helicity eigenstates with different weights determined by the differential cross section.
The distribution of the differential cross section with four helicity channels is shown in Fig.~\ref{fig4}.  ${\rm d}\sigma_{+-+-}$ and ${\rm d}\sigma_{+--+}$ present the asymmetric distributions with respect to $\theta_s$. The former produces $\mu^-$ ($\mu^+$) with a helicity state $|+\rangle$ ($|-\rangle$) located into $\theta_s<\pi/2$ and the later produces $\mu^-$ ($\mu^+$) with a helicity state $|-\rangle$ ($|+\rangle$) located into  $\theta_s>\pi/2$. Consequently, they lead to the LSP $\mu^\pm$ [see Figs.~\ref{fig4}(b) and (c)].
${\rm d}\sigma_{+-\pm\pm}$ dominate the reaction around $\theta_s=\pi/2$ and threshold energy, and produce $\mu^\pm$ with a helicity state $|+\rangle$ or $|-\rangle$ by the same weight [see Fig.~\ref{fig4}(d)], thus they contribute the purely transverse polarization~\cite{Hikasa1986}. Note that the spin-correlated term in Eq.~(\ref{diffCS}) has the non-negligible contribution to the ${\rm d}\sigma_{+-++}$ and ${\rm d}\sigma_{+---}$ channels, and subsequently to the transverse polarization.
The weights that $\mu^\pm$ possess the definite helicity eigenstates with are determined by the ratios between different cross sections of each helicity channel and ${\rm d}\bar{\sigma}_{+-}/{\rm d}\theta_s$, and the superposition of these four helicity channels with corresponding weights leads to the helicity distribution (see Fig.~\ref{fig1}).

The helicity distributions  determine the polarization characteristics of produced muons.
For the $e^{-}e^+$ collision with opposite helicity eigenstates,  the  distributions of the total polarization and its LSP and TSP components are shown in Fig.~\ref{fig5}.
$P^{(\mu)}_{\parallel}$ originates from the separated distributions of the ${\rm d}\sigma_{+-+-}$ and ${\rm d}\sigma_{+--+}$ channels, implying $P^{(\mu)}_{\parallel}=\left|\frac{{\rm d}\sigma_{+-+-}-{\rm d}\sigma_{+--+}}{{\rm d}\sigma_{+-+-}+{\rm d}\sigma_{+--+}}\right|$ and leading to the symmetric distribution of $P^{(\mu)}_{\parallel}$  with respect to $\theta_s$ [see Fig.~\ref{fig5}(b)]. While, $P^{(\mu)}_{\perp}$ originates from the same distributions of the ${\rm d}\sigma_{+-++}$ and ${\rm d}\sigma_{+---}$ channels, and the partial overlap between ${\rm d}\sigma_{+-\pm\pm}$ and ${\rm d}\sigma_{+-\pm\mp}$ leads to the non-equal weights of $\mu^\pm$ at the helicity states $|\pm\rangle$  and subsequently the partially transverse polarization [see Fig.~\ref{fig5}(c)]. The relation $P^{(\mu)}_{tot}=\sqrt{(P^{(\mu)}_{\parallel})^2+(P^{(\mu)}_{\perp})^2}$ leads to the distribution of the total polarization [see Fig.~\ref{fig5}(a)]. The MC simulation results are consistent well with the analytical calculations [see Figs.~\ref{fig5}(d)-(f)]. According to the polarization distribution, in the polarized $e^-e^+$ collisions, the LSP muons can be extracted from the region of $\theta\lesssim 0.3\pi$ or $\theta\gtrsim0.7\pi$, and the TSP muons can be extracted from the region of $0.3\pi\lesssim\theta\lesssim 0.7\pi$.

 Since the polarization of muons are transferred from the LSP electron (positron) beams with the longitudinal polarization $P^{(e)}_{\parallel}$, it is nontrivial to illustrate the polarization curve of the muon beam versus $P^{(e)}_{\parallel}$, as shown in Fig.~\ref{fig6}. Because the $\mu^-$ ($\mu^+$) helicity is symmetric with respect to $\theta_s$ (see Fig.~\ref{fig1}), $P^{(\mu)}_{\parallel}$ calculated from all angle-distributed $\mu^-$ or $\mu^+$ is vanished, and thus $P^{(\mu)}_{tot}=P^{(\mu)}_{\perp}$ increases linearly with $P^{(e)}_{\parallel}$ [see Fig.~\ref{fig6}(a)]. The LSP muons are extracted from the paraxial region where $P^{(\mu)}_{\parallel}$ dominates the polarization, and since ${\rm d}\sigma_{+-\pm\pm}/{\rm d}\theta_s$ contributes to $P^{(\mu)}_{\perp}$ at the low energy, the polarization curve of $P^{(\mu)}_{\parallel}$ is slightly bent [see Fig.~\ref{fig6}(b)]. The TSP muons are scattered around $\theta=\pi/2$, and the polarization curve of $P^{(\mu)}_{\perp}$ is distinguished between TSP electrons and positrons with anti-parallel or parallel spin direction. The anti-parallel case with $\bm{\zeta}_{+,\perp}=-\bm{\zeta}_{-,\perp}$ leads to the dominated $P^{(\mu)}_{\perp}$ with linear growth, while the parallel case with $\bm{\zeta}_{+,\perp}=\bm{\zeta}_{-,\perp}$ leads to the dominated $P^{(\mu)}_{\perp}$ with approximately flat variation [see Fig.~\ref{fig6}(c)].
 The reasons that the polarization curve of $P^{(\mu)}_{\perp}$ depends on the spin directions are that for $\bm{\zeta}_{+,\perp}=\bm{\zeta}_{-,\perp}$, the completely overlapping helicity channels ${\rm d}\sigma_{+-\pm\pm}/{\rm d}\theta_s$ are larger than  those with $\bm{\zeta}_{+,\perp}=-\bm{\zeta}_{-,\perp}$, and the asymmetric ${\rm d}\sigma_{+-\pm\mp}/{\rm d}\theta_s$ are almost non-overlapping and thus have negligible contribution to the transverse polarization around $\theta=\pi/2$. The reactions between polarized electron beam and nonpolarized positron beam (or vice versa) also produce the similar polarization curve to Fig.~\ref{fig6} but with the half-reduced yield.

\begin{table}[!t]
 \caption{In a symmetric colliding setup with a set of example parameters: charge $Q_b$, longitudinal size $l_z$ and transverse size $\sigma_{x/y}$, and repetition frequency $f_{rep}$, the produced muons with total yield $N_{tot}$, and partial yields $N_{\parallel}$  and $N_{\perp}$. The electron and positron beams with $P^{(e)}_{\parallel}=1$  have the same energy distribution as that in Fig.~\ref{fig6}.}
 \label{Tab:beam}
 \setlength{\tabcolsep}{4pt}
 \begin{center}
  \begin{tabular}{cccccc}
   \hline\hline\specialrule{0em}{4pt}{1pt}
   $Q_b$ &($l_z, \sigma_{x/y}$) [$\mu$m]& $f_{rep}$ [Hz] &$N_{tot}$&($N_{\parallel}, P^{(\mu)}_{\parallel}$)&($N_{\perp}, P^{(\mu)}_{\perp}$) \\
   \hline\specialrule{0em}{4pt}{2pt}
   5 nC &(100, 0.3)&$10^4$&$546$&(236, 0.79)&(310, 0.83)\\
   \hline\specialrule{0em}{4pt}{2pt}
   500 pC&(50, 0.3)&$10^6$&$591$&(257, 0.83)&(334, 0.83)\\
   \hline\hline
  \end{tabular}
 \end{center}
\end{table}
The muon yields produced from a symmetric colliding setup with specific beams  are shown in Tab.~\ref{Tab:beam}, where $f_{rep}$ is estimated by the colliding beams with an equivalent 500 nC charge (see the explanation in Sec.~\ref{twoB}). The results indicate that about $40\%$ LSP muon pairs ($\theta\lesssim0.3\pi$ and $\theta\gtrsim0.7\pi$) and about $60\%$ TSP muon pairs ($0.3\pi\lesssim\theta\lesssim0.7\pi$) with approximate $80\%$ polarization degree are produced in the assumed colliding setup. The higher degree of polarization with approximate $90\%$ for LSP and TSP muons can be filtered by the narrower regions of $\theta$ (see Fig. \ref{fig6}).
 Moreover, the beam-beam collision can be designed as a charge-asymmetric setup, namely, with a nC electron beam and a hundreds-of-pC positron beam, which can significantly reduce the beam disruption at the interaction point \cite{Esberg2014}. Furthermore, the emittance of produced muons can be reduced by the tunable energy asymmetry of tens of MeV between electron and positron beams. However, we find that the polarized muons can not be produced in the colliding electron and positron beams with a GeV energy asymmetry due to the vanished average spin components, thus the interaction of a positron beam and target electrons can not generate the polarized muon source \cite{Antonelli2016,Boscolo2018,Boscolo2020}.

The implement of the leptonic $e^-e^+\rightarrow\mu^-\mu^+$ muon source depends on the high-luminosity $e^-e^+$  collisions. As an advanced next-generation collider, the laser-plasma-accelerator-based linear $e^-e^+$ colliders have attract the broad interests recently due to its compact scale \cite{Schroeder2010,Schroeder2012,Nakajima2019,Shiltsev2021}. The theoretical calculations assess that, by virtue of the kHz laser pulse, the laser-driven wakefield sustains the multi-bunch beams with continuous transverse focusing to produce a nanometer beam size, which permits laser-plasma linear collider operating at very high luminosity of the order of $10^{34}$ cm$^{-2}$s$^{-1}$ \cite{Schroeder2012,Nakajima2019}.
We underline that the dense electron beams with tens of nC charge and  hundreds of MeV energy can be generated through the direct laser acceleration  \cite{wang2017,Ma6980} or laser wakefield acceleration (LWFA) \cite{otz2020,Shaw2021}. The interaction of laser-wakefield electron beams with high-$Z$ target induce the production of the positron shower with tens of nC charge and the energy exceeding 100 MeV \cite{Alejo2019,Alejo2019laser}, which could be trapped and accelerated as a high-quality beam in the laser (beam) wakefield \cite{Sahai2018,Diederichs2020,Zhou2021}.
Thus the laser-driven electron and positron beams with ultrahigh charge afford the ultrahigh-luminosity laser-plasma linear collider for the generation of $e^-e^+\rightarrow\mu^-\mu^+$ muon source.

\section{CONCLUSION}\label{four}

In summary, we investigate the production of high-flux ASP muon pairs via the  collision of laser-driven polarized electron and positron beams. We calculate the cross section of the muon pair production with arbitrary polarization and use the spin-resolved MC method to describe the polarized muon pair production. The LSP and TSP muon pairs can be produced simultaneously in a single beam-beam collision. The polarization mechanism of the muon pair production is clarified through the differential cross section with different helicity channels.
Our calculations indicate that based on the current platform of laser-plasma acceleration, it is potential to generate LSP electron and positron beams with the charge from hundreds of pC to tens of nC, and thus could produce ASP muon pairs via the ultrahigh-luminosity laser-plasma collider scheme. Our proposed  leptonic muon source could be applicable in broad areas, such as laser-plasma, nuclear and high-energy particle and condensed matter physics.

\section{ACKNOWLEDGEMENT}
This work is supported by the National Natural Science Foundation of China (Grants Nos. 12022506, 11874295, 11875219, 11655003, 11905169, 12105217), the China Postdoctoral Science Foundation (Grant No. 2020M683447), the Innovation Project
of IHEP (542017IHEPZZBS11820, 542018IHEPZZBS12427), and the CAS Center for Excellence in Particle Physics
(CCEPP).

\appendix
\section{Numerical simulation method}\label{appA}
 In the simulations, the realistic $e^-$ and $e^+$ (not macroparticles) are randomized with energy and spatial distributions. Following the algorithm in \cite{Gaudio2020}, we consider a collision cell containing $N_{-}$ electrons in the beam (1) and $N_{+}$ positrons in the beam (2), the maximum probability of a  collision within an unit time $\Delta \tau$ and an unit volume $\Delta V$ is
\begin{equation}
P_{max}=2\sigma_Tc\Delta \tau/\Delta V,
\end{equation}
where $\sigma_T$ is the Thomson cross section and used as the reference cross section. At each colliding time step, the maximal number of $e^-$ and $e^+$ in a cell that are probable to scatter is $N_{max}=P_{max}N_{-}N_{+}$. After randomly sorting the particles of each beam in a cell, the $e^-$ and $e^+$ used to be paired for collision are selected from the first $N_{max}$ particles in each sorted $e^-$ or $e^+$ list. The event probability is given by
\begin{equation}
P^{i,j}=(\bar{\sigma}_{tot}c \Delta \tau/\Delta V)/P_{max}.
\end{equation}
The collision of a paired $e^{\pm}$ is admitted to the $\mu^{\pm}$ pair production process based on a rejection method, i.e., for a random number $R_0$, accept if $P^{i,j}>R_0$.

\begin{figure}[!h]
	\setlength{\abovecaptionskip}{-0.2cm}
	\centering\includegraphics[width=1\linewidth]{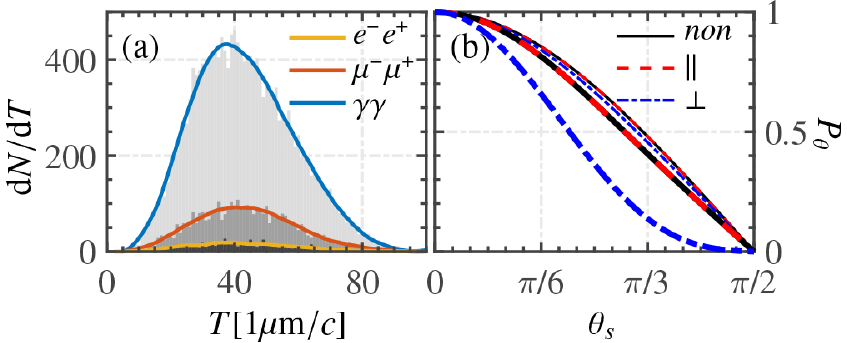}
	\caption{(a) Reaction rates of three different processes in the collision of $e^-$ and $e^+$ beams. (b) $P_\theta=\bar{\sigma}_\theta/\bar{\sigma}_{tot}$ vs $\theta_s$ for nonpolarized ($non$), LSP ($\parallel$) and TSP ($\perp$) $e^\pm$, calculated from Eqs.~(\ref{intCS}) and (\ref{sigtheta}). The thin and thick lines correspond to  $E_{cm}$ = 125 MeV and 1 GeV, respectively.}
	\label{fig7}
\end{figure}

If we consider the competitive one-order QED processes in each $e^-e^+$ collision:  $e^-e^+\rightarrow\gamma\gamma$, $e^-e^+\rightarrow\mu^-\mu^+$ and $e^-e^+\rightarrow e^-e^+$, the event is determined by the selection method.
We define $P_{\gamma\gamma}=\bar\sigma_{e^-e^+\rightarrow\gamma\gamma}/2\sigma_T$, $P_{\mu^-\mu^+}=\bar\sigma_{e^-e^+\rightarrow\mu^-\mu^+}/2\sigma_T$ and $P_{e^-e^+}=\bar\sigma_{e^-e^+\rightarrow e^-e^+}/2\sigma_T$.
Randomize a uniform number $R_0$ between 0 and 1 and execute the following selections:
\begin{itemize}
	\item $R_0\in(0,P_{\gamma\gamma})$ leads to the annihilation process;
	\item $R_0\in(P_{\gamma\gamma},P_{\mu^-\mu^+}+P_{\gamma\gamma})$ leads to the muon pair production process;
	\item  $R_0\in(P_{\mu^-\mu^+}+P_{\gamma\gamma},P_{e^-e^+}+P_{\mu^-\mu^+}+P_{\gamma\gamma})$ leads to the elastic scattering process;
	\item $R_0\in(P_{e^-e^+}+P_{\mu^-\mu^+}+P_{\gamma\gamma},1)$ leads to no any reaction.
\end{itemize}
The production yield of these three processes in the collision of $e^-$ and $e^+$ beams is shown in Fig.~\ref{fig7}(a).

With the definition $P_{\theta}=\bar{\sigma}_{\theta}/\bar{\sigma}_{tot}$, the distribution of $P_{\theta}$ implies that the produced pair is scattered into ${\rm d}\theta_s$ with a probability ${\rm d}P_{\theta}$, as shown in Fig.~\ref{fig7} (b). Thus, $\theta_s$ can be determined by solving the equation $P_\theta(\theta_s)=R_1$, where $R_1$ is a uniform random number between -1 and 1. After the determination of $\theta_s$, the momenta of $\mu^\pm$ in the c.m. frame can be determined according to the momentum relations presented in Fig.~\ref{fig2}. The energy and momentum of muons in the laboratory frame are calculated by the Lorentz transformation
\begin{eqnarray}\label{gamma}
\gamma'_{\pm}&=&\gamma'\gamma_{cm}(1\mp\bm{\beta}^2_{cm}\bm{\beta}')\cos{\theta_u}~,\\
\bm{p}'_{\pm}&=&\left(\bm{p}'\mp\frac{(\gamma_{cm}-1)}{\beta_{cm}^2}\beta'\gamma'\cos{\theta_u}\bm{\beta}_{cm}\right)\mp\frac{\gamma_{cm}}{\beta'}\bm{\beta}_{cm}~,
\end{eqnarray}
where $\gamma_{cm}=1/\sqrt{1-\beta_{cm}^2}$, and $\gamma'\beta'=|\bm{p}'|$.  $\theta_u$ is the angle between $\bm{\beta}_{cm}$ and $\bm{p}'$.

In the MC procedure, the projection of the spin axis $\bm{\zeta}'^{(f)}_{\pm}$ onto the spin state $\pm\bm{\zeta}_\pm^{(d)}$ of a detector is determined by a random number $R_2$ between 0 and 1:
\begin{itemize}
	\item $R_2\in(0,W^{\uparrow\uparrow})$ results in $+\zeta_{+,i}^{(d)},~+\zeta_{-,i}^{(d)}$;
	\item $R_2\in(W^{\uparrow\uparrow},W^{\downarrow\downarrow}+W^{\uparrow\uparrow})$ results in $-\zeta_{+,i}^{(d)},~-\zeta_{-,i}^{(d)}$;
	\item $R_2\in(W^{\downarrow\downarrow}+W^{\uparrow\uparrow},W^{\uparrow\downarrow}+W^{\downarrow\downarrow}+W^{\uparrow\uparrow})$ results in $+\zeta_{+,i}^{(d)},~-\zeta_{-,i}^{(d)}$;
	\item  $R_2\in(W^{\uparrow\downarrow}+W^{\downarrow\downarrow}+W^{\uparrow\uparrow},1)$ results in $-\zeta_{+,i}^{(d)},~+\zeta_{-,i}^{(d)}$.
\end{itemize}

Finally, the execution of the simulation code is performed as follows:
\begin{enumerate}
	\item Initializing the $e^{\pm}$ beams with an energy distribution, angle divergence and mean spin.
	\item Griding the collision region into 3D cells at simulation time $T_1$.
	\item Paring the colliding $e^{\pm}$ inside the first cell $\Delta x_1\Delta y_1\Delta z_1$ using the no-time-count method, and transforming each pair from the laboratory from to the c.m. frame.
	\item Randomizing a number $R_0$ and using the rejection method to generate $\mu^{\pm}$ pairs inside the cell. If $P^{i,j}$ is too large (say, $>0.1$), divide the interval $\Delta \tau$ (and $P^{i,j}$) by an integer $N_{div}$, and repeat the following procedure $N_{div}$ times.
	\item Randomizing another number $R_1$ between -1 and 1 to determine the scattering angle $\theta_s$ and hence the momenta of created pair in the c.m. frame.
	\item Calculating Eqs.~(\ref{zeta}) and (\ref{polarization vector}) by the determined momenta, and randomizing a number $R_2$ to determine the observable spin state.
	\item Going back to the step 3 and undergoing the next paired collision of $e^{\pm}$.
\end{enumerate}

\section{The expressions of the coefficients in Eq. (\ref{diffCS})} \label{appB}

The Mandelstam invariants in the process of $e^-e^+\rightarrow \mu^-\mu^+$ are defined as
\begin{widetext}
\begin{eqnarray}
s&=(p_-+p_+)^2=(p'_-+p'_+)^2;~~~\tilde{s}=s/m_e^2 , \nonumber \\
t&=(p_--p'_-)^2=(p'_+-p_+)^2;~~~\tilde{t}=t/m_e^2 , \nonumber \\
u&=(p_--p'_+)^2=(p'_--p_+)^2;~~~\tilde{u}=u/m_e^2 .
\end{eqnarray}
\end{widetext}

 The final-state spin irrelevant term is
\begin{widetext}
\begin{eqnarray}
F&=\frac{1}{(\tilde{s}-4)^2}8(-2m^4(\tilde{s}-4)^2(\zeta _{-,1}\zeta _{+,1}+\zeta _{-,2}\zeta _{+,2}-\zeta _{-,3}\zeta _{+,3}-1)+2(-(\tilde{s}-4)\zeta _{-,2}\zeta _{+,3}\tilde{t}\sqrt{-\tilde{s}(4 m^2 (\tilde{s}-4)-\tilde{s}^2+4 \tilde{s}+(\tilde{t}-\tilde{u})^2)}   \nonumber  \\
&+(\tilde{s}-4)\zeta _{-,2}\zeta _{+,3}\tilde{u}\sqrt{-\tilde{s}(4 m^2 (\tilde{s}-4)-\tilde{s}^2+4 \tilde{s}+(\tilde{t}-\tilde{u})^2)}+(\tilde{s}-4)\zeta _{+,2}\zeta _{-,3}\tilde{t}\sqrt{-\tilde{s}(4 m^2 (\tilde{s}-4)-\tilde{s}^2+4 \tilde{s}+(\tilde{t}-\tilde{u})^2)}  \nonumber \\
&-(\tilde{s}-4)\zeta _{+,2}\zeta _{-,3}\tilde{u}\sqrt{-\tilde{s}(4 m^2 (\tilde{s}-4)-\tilde{s}^2+4 \tilde{s}+(\tilde{t}-\tilde{u})^2)}-16\zeta _{-,1}\zeta _{+,1}\tilde{t}\tilde{u}+16\zeta _{-,1}\zeta _{+,1}\tilde{t}+16\zeta _{-,1}\zeta _{+,1}\tilde{u}-16\zeta _{-,1} \zeta _{+,1}   \nonumber  \\
&-16\zeta _{-,2} \zeta _{+,2} \tilde{t} \tilde{u}+16 \zeta _{-,2} \zeta _{+,2} \tilde{t}+16 \zeta _{-,2} \zeta _{+,2} \tilde{u}-16 \zeta _{-,2} \zeta _{+,2}+8 \zeta _{-,3} \zeta _{+,3} \tilde{t}^2-16 \zeta _{-,3} \zeta _{+,3} \tilde{t}+8 \zeta _{-,3} \zeta _{+,3} \tilde{u}^2-16 \zeta _{-,3} \zeta _{+,3} \tilde{u} +16 \zeta _{-,3} \zeta _{+,3}   \nonumber  \\
&+8 \tilde{t}^2-16 \tilde{t}+8 \tilde{u}^2-16 \tilde{u}+16)+2 m^2 (\tilde{s}-4)^2(\tilde{s}(\zeta _{-,1} \zeta _{+,1}-\zeta _{-,2} \zeta _{+,2}-\zeta _{-,3} \zeta _{+,3}+1)+(\tilde{t}+\tilde{u}-2)(\zeta _{-,1} \zeta _{+,1}+\zeta _{-,2} \zeta _{+,2}  \nonumber  \\
&-\zeta _{-,3} \zeta _{+,3}-1))+2 \tilde{s}^3 (\zeta _{-,1} \zeta _{+,1}+\zeta _{-,2} \zeta _{+,2}+\zeta _{-,3}\zeta _{+,3}+1)-\tilde{s}^2(2\tilde{t}(\zeta _{-,1}\zeta _{+,1}(\tilde{u}-1)-\zeta _{-,2}\zeta _{+,2}-\zeta _{-,3}\zeta _{+,3}\tilde{u}+\zeta _{-,3}\zeta _{+,3}+1) \nonumber \\
&-2\tilde{u}(\zeta _{-,1}\zeta _{+,1}+\zeta _{-,2}\zeta _{+,2}-\zeta _{-,3}\zeta _{+,3}-1)+2(9\zeta _{-,1}\zeta _{+,1}+9\zeta _{-,2}\zeta _{+,2}+7\zeta _{-,3}\zeta _{+,3}+7)+\tilde{t}^2(\zeta _{-,2}\zeta _{+,2}-1)+\tilde{u}^2(\zeta _{-,2}\zeta _{+,2}-1)) \nonumber  \\
&+4\tilde{s}(2\tilde{t}(2\zeta _{-,1}\zeta _{+,1}(\tilde{u}-1)+\zeta _{-,2}\zeta _{+,2}(\tilde{u}-2)-\zeta _{-,3}\zeta _{+,3}\tilde{u}+2\zeta _{-,3}\zeta _{+,3}+2)-4\tilde{u}(\zeta _{-,1}\zeta _{+,1}+\zeta _{-,2}\zeta _{+,2}-\zeta _{-,3}\zeta _{+,3}-1) \nonumber \\
&+4(3\zeta _{-,1}\zeta _{+,1}+3\zeta _{-,2}\zeta _{+,2}+\zeta _{-,3}\zeta _{+,3}+1)+\tilde{t}^2(\zeta _{-,2}\zeta _{+,2} -\zeta _{-,3}\zeta _{+,3}-2)+\tilde{u}^2(\zeta _{-,2}\zeta _{+,2}-\zeta _{-,3}\zeta _{+,3}-2))) ~.
\end{eqnarray}
\end{widetext}

The coefficients of $\zeta'_{-,i}$ are
\begin{widetext}
\begin{subequations}
	\begin{align}
	G^-_{1} &= -32m(\zeta _{-,1}+\zeta _{+,1})(2m^2-\tilde{t}-\tilde{u}+2) ~, \\
	G^-_{2} &= -\frac{1}{m (\tilde{s}-4)(4 m^2-\tilde{s})}8\sqrt{m^2 \tilde{s} (\tilde{s}-4 m^2)}(2 m^2-\tilde{t}-\tilde{u}+2)(-2(\zeta _{-,3}+\zeta _{+,3})\sqrt{-(4 m^2 (\tilde{s}-4)-\tilde{s}^2+4 \tilde{s}+(\tilde{t}-\tilde{u})^2)}  \nonumber\\
	&\times\sqrt{(\tilde{s}-4)}+\sqrt{(\tilde{s}-4) \tilde{s}}\tilde{t}(\zeta _{-,2}-\zeta _{+,2})-\sqrt{(\tilde{s}-4) \tilde{s}}\tilde{u}(\zeta _{-,2}-\zeta _{+,2})) ~,\\
	G^-_{3} &= (16m(2 m^2-\tilde{t}-\tilde{u}+2)((\tilde{t}-\tilde{u})(-2(\zeta _{-,3}+\zeta _{+,3})\sqrt{-(\tilde{s}-4) (4 m^2 (\tilde{s}-4)-\tilde{s}^2+4 \tilde{s}+(\tilde{t}-\tilde{u})^2)}+\sqrt{(\tilde{s}-4) \tilde{s}}\tilde{t}(\zeta _{-,2}\nonumber\\
	&-\zeta _{+,2}) -\sqrt{(\tilde{s}-4) \tilde{s}}\tilde{u}(\zeta _{-,2}-\zeta _{+,2}))+4m^2(\tilde{s}-4)\sqrt{(\tilde{s}-4) \tilde{s}}(\zeta _{-,2}-\zeta _{+,2})-\tilde{s}^2\sqrt{(\tilde{s}-4) \tilde{s}}(\zeta _{-,2}-\zeta _{+,2})  \nonumber\\
	&+4\tilde{s}\sqrt{(\tilde{s}-4) \tilde{s}}(\zeta _{-,2}-\zeta _{+,2})))/((\tilde{s}-4)\sqrt{(\tilde{s}-4)(4 m^2-\tilde{s})} \sqrt{\frac{4 m^2 (\tilde{s}-4)-\tilde{s}^2+4 \tilde{s}+(\tilde{t}-\tilde{u})^2}{\tilde{s}-4}} ) ~.
	\end{align}
\end{subequations}
\end{widetext}

The coefficients of $\zeta'_{+,i}$ are
\begin{widetext}
	\begin{subequations}
		\begin{align}
	G^+_{1} &= -32m(\zeta _{-,1}+\zeta _{+,1})(2m^2-\tilde{t}-\tilde{u}+2)~, \\
	G^+_{2} &= \frac{1}{m (\tilde{s}-4)(4 m^2-\tilde{s})}8\sqrt{m^2 \tilde{s} (\tilde{s}-4 m^2)}(2 m^2-\tilde{t}-\tilde{u}+2)(-2(\zeta _{-,3}+\zeta _{+,3})\sqrt{-(4 m^2 (\tilde{s}-4)-\tilde{s}^2+4 \tilde{s}+(\tilde{t}-\tilde{u})^2)}  \nonumber\\
	&\times\sqrt{(\tilde{s}-4)}+\sqrt{(\tilde{s}-4) \tilde{s}}\tilde{t}(\zeta _{-,2}-\zeta _{+,2})-\sqrt{(\tilde{s}-4) \tilde{s}}\tilde{u}(\zeta _{-,2}-\zeta _{+,2})) ~,\\
	G^+_{3} &= (16m(2 m^2-\tilde{t}-\tilde{u}+2)((\tilde{t}-\tilde{u})(-2(\zeta _{-,3}+\zeta _{+,3})\sqrt{-(\tilde{s}-4) (4 m^2 (\tilde{s}-4)-\tilde{s}^2+4 \tilde{s}+(\tilde{t}-\tilde{u})^2)}+\sqrt{(\tilde{s}-4) \tilde{s}}\tilde{t}(\zeta _{-,2}  \nonumber\\
	&-\zeta _{+,2})-\sqrt{(\tilde{s}-4) \tilde{s}}\tilde{u}(\zeta _{-,2}-\zeta _{+,2}))+4m^2(\tilde{s}-4)\sqrt{(\tilde{s}-4) \tilde{s}}(\zeta _{-,2}-\zeta _{+,2})-\tilde{s}^2\sqrt{(\tilde{s}-4) \tilde{s}}(\zeta _{-,2}-\zeta _{+,2})  \nonumber\\
	&+4\tilde{s}\sqrt{(\tilde{s}-4) \tilde{s}}(\zeta _{-,2}-\zeta _{+,2})))/((\tilde{s}-4)\sqrt{(\tilde{s}-4)(4 m^2-\tilde{s})} \sqrt{\frac{4 m^2 (\tilde{s}-4)-\tilde{s}^2+4 \tilde{s}+(\tilde{t}-\tilde{u})^2}{\tilde{s}-4}} ) ~.
	\end{align}
\end{subequations}
\end{widetext}

The coefficients of $\zeta'_{-,i}\zeta'_{+,j}$ are
\begin{widetext}
	\begin{subequations}
		\begin{align}
		H_{11}&=\frac{1}{(\tilde{s}-4)^2 \tilde{s}}8(2 m^4 (\tilde{s}-4)^2 (\tilde{s} (\zeta _{-,1} \zeta _{+,1}-\zeta _{-,2} \zeta _{+,2}-\zeta _{-,3} \zeta _{+,3} +1)+4 (\zeta _{-,1} \zeta _{+,1}+\zeta _{-,2} \zeta _{+,2}+\zeta _{-,3} \zeta _{+,3}+1)) -2 \tilde{s} (\tilde{t} (\sqrt{\tilde{s}}  \nonumber\\
		&\times(\tilde{s}-4)\zeta _{-,2} \zeta _{+,3}\sqrt{-(4 m^2 (\tilde{s}-4)-\tilde{s}^2+4 \tilde{s}+(\tilde{t}-\tilde{u})^2)}-(\tilde{s}-4)\zeta _{+,2} \zeta _{-,3} \sqrt{-\tilde{s}(4 m^2 (\tilde{s}-4)-\tilde{s}^2+4 \tilde{s}+(\tilde{t}-\tilde{u})^2)}   \nonumber\\
		&+16 \zeta _{-,1} \zeta _{+,1} \tilde{u}-16 \zeta _{-,1} \zeta _{+,1}+32 \zeta _{-,2} \zeta _{+,2} \tilde{u}-48 \zeta _{-,2} \zeta _{+,2}+32 \zeta _{-,3} \zeta _{+,3} \tilde{u}-48 \zeta _{-,3} \zeta _{+,3}+16 \tilde{u}-16)-\tilde{u} (\sqrt{(\tilde{s}-4) \tilde{s}}  \nonumber\\
		&\times\zeta _{-,2} \zeta _{+,3} \sqrt{-(\tilde{s}-4) (4 m^2 (\tilde{s}-4)-\tilde{s}^2+4 \tilde{s}+(\tilde{t}-\tilde{u})^2)}-(\tilde{s}-4)\zeta _{+,2} \zeta _{-,3} \sqrt{-\tilde{s}(4 m^2 (\tilde{s}-4)-\tilde{s}^2+4 \tilde{s}+(\tilde{t}-\tilde{u})^2)}         \nonumber\\
		&+16 \zeta _{-,1} \zeta _{+,1}+48 \zeta _{-,2} \zeta _{+,2}+48 \zeta _{-,3} \zeta _{+,3}+16)+16 (\zeta _{-,1} \zeta _{+,1}+3 \zeta _{-,2} \zeta _{+,2}+3 \zeta _{-,3} \zeta _{+,3}+1)+8 \tilde{t}^2 (\zeta _{-,2} \zeta _{+,2}+\zeta _{-,3} \zeta _{+,3})  \nonumber\\
		&+8 \tilde{u}^2 (\zeta _{-,2} \zeta _{+,2}+\zeta _{-,3} \zeta _{+,3}))+2 m^2 (\tilde{s}-4)^2 (\tilde{s}^2 (\zeta _{-,1} \zeta _{+,1}-\zeta _{-,2} \zeta _{+,2}-\zeta _{-,3} \zeta _{+,3}+1)-\tilde{s} (\tilde{t}+\tilde{u}-2) (\zeta _{-,1} \zeta _{+,1}-\zeta _{-,2} \zeta _{+,2} \nonumber\\
		&-\zeta _{-,3} \zeta _{+,3}+1)-4 (\tilde{t}+\tilde{u}-2) (\zeta _{-,1} \zeta _{+,1}+\zeta _{-,2} \zeta _{+,2}+\zeta _{-,3} \zeta _{+,3}+1))+\tilde{s}^5 (\zeta _{-,2} \zeta _{+,2}-1)+\tilde{s}^4 (8-8 \zeta _{-,2} \zeta _{+,2})+\tilde{s}^3 (\tilde{t}^2 (\zeta _{-,1} \zeta _{+,1}    \nonumber\\
		&-\zeta _{-,2} \zeta _{+,2}-\zeta _{-,3} \zeta _{+,3}+1)+\tilde{t} (-2 \zeta _{-,1} \zeta _{+,1}+2 \zeta _{-,2} \zeta _{+,2}+2 \zeta _{-,3} \zeta _{+,3}-2)+\tilde{u}^2 (\zeta _{-,1} \zeta _{+,1}-\zeta _{-,2} \zeta _{+,2}-\zeta _{-,3} \zeta _{+,3}+1)  \nonumber\\
		&+\tilde{u} (-2 \zeta _{-,1} \zeta _{+,1}+2 \zeta _{-,2} \zeta _{+,2}+2 \zeta _{-,3} \zeta _{+,3}-2)+2 (\zeta _{-,1} \zeta _{+,1}+7 \zeta _{-,2} \zeta _{+,2}-\zeta _{-,3} \zeta _{+,3}-7))+\tilde{s}^2 (\tilde{t}^2 (-6 \zeta _{-,1} \zeta _{+,1}+6 \zeta _{-,2} \zeta _{+,2}  \nonumber\\
		&+6 \zeta _{-,3} \zeta _{+,3}-6)+4 \tilde{t} (\tilde{u} (\zeta _{-,1} \zeta _{+,1}+3 \zeta _{-,2} \zeta _{+,2}+3 \zeta _{-,3} \zeta _{+,3}+1)+2 \zeta _{-,1} \zeta _{+,1}-6 \zeta _{-,2} \zeta _{+,2}-6 \zeta _{-,3} \zeta _{+,3}+2)+\tilde{u}^2 (-6 \zeta _{-,1} \zeta _{+,1}   \nonumber\\
		&+6 \zeta _{-,2} \zeta _{+,2}+6 \zeta _{-,3} \zeta_{+,3}-6)+8 \tilde{u} (\zeta _{-,1} \zeta _{+,1}-3 \zeta_{-,2} \zeta _{+,2}-3 \zeta _{-,3} \zeta_{+,3}+1)-8 (\zeta _{-,1} \zeta _{+,1}-3 \zeta_{-,2} \zeta _{+,2}-3 \zeta _{-,3} \zeta_{+,3}+1))  \nonumber\\
		&+32 (\tilde{t}+\tilde{u}-2)^2 (\zeta_{-,1} \zeta_{+,1}+\zeta_{-,2} \zeta_{+,2}+\zeta_{-,3} \zeta_{+,3}+1)) 	~,\\
		H_{12}&=-\frac{1}{((\tilde{s}-4) \tilde{s})^{3/2} \sqrt{m^2 \tilde{s} (\tilde{s}-4 m^2)}}32m^2\tilde{s}^3(-2m^2 \tilde{s} \zeta _{-,1} \zeta _{+,2}-(\tilde{s}-4) \zeta _{+,1} \zeta _{-,2}( -2 m^2+\tilde{s}+2 \tilde{t}-2)+8 m^2 \zeta _{-,1} \zeta _{+,2}  \nonumber\\
		&+(\tilde{s}-4)\zeta _{-,1} \zeta _{+,3} \sqrt{-\tilde{s}(m^4-2 m^2 (\tilde{t}+1)+(\tilde{s}-2) \tilde{t}+\tilde{t}^2+1)}+(\tilde{s}-4)\zeta _{+,1} \zeta _{-,3} \sqrt{-\tilde{s}(m^4-2 m^2 (\tilde{t}+1)+(\tilde{s}-2) \tilde{t}+\tilde{t}^2+1)}   \nonumber\\
		&+\tilde{s}^2 \zeta _{-,1}\zeta _{+,2}+2\tilde{s}\zeta _{-,1}\zeta _{+,2}\tilde{t}-6\tilde{s}\zeta _{-,1}\zeta _{+,2}-8\zeta _{-,1}\zeta _{+,2}\tilde{t}+8\zeta _{-,1}\zeta _{+,2}) ~,\\
		H_{13}&=(8\tilde{s}(\tilde{s}^2\sqrt{-(\tilde{s}-4)(m^4-2 m^2 (\tilde{t}+1)+(\tilde{s}-2) \tilde{t}+\tilde{t}^2+1)}(\zeta _{-,1}\zeta _{+,3}+\zeta _{+,1}\zeta _{-,3})-2\tilde{s}(-\zeta _{-,1}\tilde{t}(\zeta _{+,3}\sqrt{(\tilde{s}-4)} \nonumber\\
		&\times\sqrt{-(m^4-2 m^2 (\tilde{t}+1)+(\tilde{s}-2) \tilde{t}+\tilde{t}^2+1)}+2\sqrt{(\tilde{s}-4) \tilde{s}}\zeta _{+,2})+(m^2+1)\zeta _{-,1}\zeta _{+,3}\sqrt{(\tilde{s}-4)}  \nonumber\\
		&\times\sqrt{-(m^4-2 m^2 (\tilde{t}+1)+(\tilde{s}-2) \tilde{t}+\tilde{t}^2+1)}+\zeta _{+,1}\zeta _{-,3}(m^2-\tilde{t}+1)\sqrt{-(\tilde{s}-4) (m^4-2 m^2 (\tilde{t}+1)+(\tilde{s}-2) \tilde{t}+\tilde{t}^2+1)}) \nonumber\\
		&+4\sqrt{(\tilde{s}-4) \tilde{s}}\zeta _{-,1}\zeta _{+,2}(m^4-2 m^2 (\tilde{t}+1)+(\tilde{t}-1)^2)-4\sqrt{(\tilde{s}-4) \tilde{s}}\zeta _{+,1}\zeta _{-,2}(m^4-2 m^2 (\tilde{t}+1)+(\tilde{s}-2) \tilde{t}+\tilde{t}^2  \nonumber\\
		&+1)))/((\tilde{s}-4) \sqrt{m^4-2 m^2 (\tilde{t}+1)+(\tilde{s}-2) \tilde{t}+\tilde{t}^2+1}\sqrt{(4 m^2-\tilde{s})}) ~,\\
		H_{21}&=\frac{1}{((\tilde{s}-4) \tilde{s})^{3/2} \sqrt{m^2 \tilde{s} (\tilde{s}-4 m^2)}}32m^2\tilde{s}^3(-2m^2 \tilde{s} \zeta _{-,1} \zeta _{+,2}-(\tilde{s}-4) \zeta _{+,1} \zeta _{-,2}( -2 m^2+\tilde{s}+2 \tilde{t}-2)+8 m^2 \zeta _{-,1} \zeta _{+,2}  \nonumber\\
		&+(\tilde{s}-4)\zeta _{-,1} \zeta _{+,3} \sqrt{-\tilde{s}(m^4-2 m^2 (\tilde{t}+1)+(\tilde{s}-2) \tilde{t}+\tilde{t}^2+1)}+(\tilde{s}-4)\zeta _{+,1} \zeta _{-,3} \sqrt{-\tilde{s}(m^4-2 m^2 (\tilde{t}+1)+(\tilde{s}-2) \tilde{t}+\tilde{t}^2+1)}   \nonumber\\
		&+\tilde{s}^2 \zeta _{-,1}\zeta _{+,2}+2\tilde{s}\zeta _{-,1}\zeta _{+,2}\tilde{t}-6\tilde{s}\zeta _{-,1}\zeta _{+,2}-8\zeta _{-,1}\zeta _{+,2}\tilde{t}+8\zeta _{-,1}\zeta _{+,2}) ~,\\	
			H_{22}&=\frac{1}{(\tilde{s}-4)^2 (4 m^2-\tilde{s})} 8(8 (\tilde{s}-4)^2(\zeta _{-,1} \zeta _{+,1}+\zeta _{-,2} \zeta _{+,2}-\zeta _{-,3} \zeta _{+,3}-1) m^6+2 (\tilde{s}-4)^2 (\tilde{s} (3 \zeta _{-,1} \zeta _{+,1}-5 \zeta _{-,2} \zeta _{+,2}-3 \zeta _{-,3} \zeta _{+,3}+5) \nonumber\\
		&-4 (\tilde{t}+\tilde{u}-2) (\zeta _{-,1} \zeta _{+,1}+\zeta _{-,2} \zeta _{+,2}-\zeta _{-,3} \zeta _{+,3}-1)) m^4-2 ((\zeta _{-,1} \zeta _{+,1}-\zeta _{-,2} \zeta _{+,2}-\zeta _{-,3} \zeta _{+,3}+1) \tilde{s}^4 +(-\zeta _{-,1} \zeta _{+,1} \tilde{u}-\zeta _{-,2} \zeta _{+,2} \tilde{u} \nonumber\\
		&+\zeta _{-,3} \zeta _{+,3} \tilde{u}+\tilde{u}-2 \zeta _{-,1} \zeta _{+,1}+14 \zeta _{-,2} \zeta _{+,2}+10 \zeta _{-,3} \zeta _{+,3}+\tilde{t} (-\zeta _{-,1} \zeta _{+,1}-\zeta _{-,2} \zeta _{+,2}+\zeta _{-,3} \zeta _{+,3}+1)-6) \tilde{s}^3-2 ((\zeta _{-,1} \zeta _{+,1}-\zeta _{-,3} \zeta _{+,3}) \tilde{t}^2  \nonumber\\
		&+2 (-3 \zeta _{-,1} \zeta _{+,1}-3 \zeta _{-,2} \zeta _{+,2}+\tilde{u} (\zeta _{-,2} \zeta _{+,2}-1)+3 \zeta _{-,3} \zeta _{+,3}+3) \tilde{t}+\tilde{u}^2 (\zeta _{-,1} \zeta _{+,1}-\zeta _{-,3} \zeta _{+,3})-6 \tilde{u} (\zeta _{-,1} \zeta _{+,1}+\zeta _{-,2} \zeta _{+,2}-\zeta _{-,3} \zeta _{+,3}-1) \nonumber\\
		&+2 (9 \zeta _{-,1} \zeta _{+,1}+17 \zeta _{-,2} \zeta _{+,2}+7 \zeta _{-,3} \zeta _{+,3}-1)) \tilde{s}^2+8 ((2 \zeta _{-,1} \zeta _{+,1}+\zeta _{-,2} \zeta _{+,2}-\zeta _{-,3} \zeta _{+,3}) \tilde{t}^2 +2 (-3 \zeta _{-,1} \zeta _{+,1}-3 \zeta _{-,2} \zeta _{+,2}+3 \zeta _{-,3} \zeta _{+,3} \nonumber\\
		&+\tilde{u} (\zeta _{-,2} \zeta _{+,2}-\zeta _{-,3} \zeta _{+,3}-2)+3) \tilde{t}+16 (\zeta _{-,1} \zeta _{+,1}+\zeta _{-,2} \zeta _{+,2})-6 \tilde{u} (\zeta _{-,1} \zeta _{+,1}+\zeta _{-,2} \zeta _{+,2}-\zeta _{-,3} \zeta _{+,3}-1)+\tilde{u}^2 (2 \zeta _{-,1} \zeta _{+,1}+\zeta _{-,2} \zeta _{+,2} \nonumber\\
		&-\zeta _{-,3} \zeta _{+,3})) \tilde{s}-4 (8 \zeta _{-,1} \zeta _{+,1} \tilde{t}^2+8 \zeta _{-,2} \zeta _{+,2} \tilde{t}^2-16 \tilde{t} \tilde{u}-16 \zeta _{-,1} \zeta _{+,1} \tilde{t}-16 \zeta _{-,2} \zeta _{+,2} \tilde{t}+(\tilde{s}-4)\sqrt{-\tilde{s}(4 m^2 (\tilde{s}-4)-\tilde{s}^2+4 \tilde{s}+(\tilde{t}-\tilde{u})^2)}  \nonumber\\
		&\times\zeta_{-,3}\zeta_{+,2}\tilde{t}-(\tilde{s}-4)\sqrt{-\tilde{s}(4 m^2 (\tilde{s}-4)-\tilde{s}^2+4 \tilde{s}+(\tilde{t}-\tilde{u})^2)} \zeta _{-,2} \zeta _{+,3} \tilde{t}-16 \tilde{u} \zeta _{-,3} \zeta _{+,3} \tilde{t}+16 \zeta _{-,3} \zeta _{+,3} \tilde{t}+16 \tilde{t}+16 \tilde{u}+8 \tilde{u}^2 \zeta _{-,1} \zeta _{+,1} \nonumber\\
		&-16 \tilde{u} \zeta _{-,1} \zeta _{+,1}+16 \zeta _{-,1} \zeta _{+,1}+8 \tilde{u}^2 \zeta _{-,2} \zeta _{+,2}-16 \tilde{u} \zeta _{-,2} \zeta _{+,2}+16 \zeta _{-,2} \zeta _{+,2}-\sqrt{-\tilde{s}(4 m^2 (\tilde{s}-4)-\tilde{s}^2+4 \tilde{s}+(\tilde{t}-\tilde{u})^2)}(\tilde{s}-4)\tilde{u} \zeta _{-,3} \zeta _{+,2} \nonumber\\
		&+ \sqrt{-\tilde{s}(4 m^2 (\tilde{s}-4)-\tilde{s}^2+4 \tilde{s}+(\tilde{t}-\tilde{u})^2)} (\tilde{s}-4)\tilde{u} \zeta _{-,2} \zeta _{+,3}+16 \tilde{u} \zeta _{-,3} \zeta _{+,3}-16 \zeta _{-,3} \zeta _{+,3}-16)) m^2+\tilde{s}(2 (\zeta _{-,1} \zeta _{+,1}+\zeta _{-,2} \zeta _{+,2} \nonumber\\
		&+\zeta _{-,3} \zeta _{+,3} +1) \tilde{s}^3-((\zeta _{-,2} \zeta _{+,2} -1) \tilde{t}^2+2((\tilde{u}-1) \zeta _{-,1} \zeta _{+,1}-\zeta _{-,2} \zeta _{+,2}-\tilde{u} \zeta _{-,3} \zeta _{+,3}+\zeta _{-,3} \zeta _{+,3}+1) \tilde{t}+\tilde{u}^2(\zeta _{-,2} \zeta _{+,2}-1)-2 \tilde{u} (\zeta _{-,1} \zeta _{+,1} \nonumber\\
		&+\zeta _{-,2} \zeta _{+,2}-\zeta _{-,3} \zeta _{+,3} -1)+2 (9 \zeta _{-,1} \zeta _{+,1}+9 \zeta _{-,2} \zeta _{+,2}+7 \zeta _{-,3} \zeta _{+,3}+7)) \tilde{s}^2+4 ((\zeta _{-,2} \zeta _{+,2}-\zeta _{-,3} \zeta _{+,3}-2) \tilde{t}^2+2 (2 (\tilde{u}-1) \zeta _{-,1} \zeta _{+,1}  \nonumber\\
		&+(\tilde{u}-2) \zeta _{-,2} \zeta _{+,2}-\tilde{u} \zeta _{-,3} \zeta _{+,3}+2 \zeta _{-,3} \zeta _{+,3}+2) \tilde{t}+\tilde{u}^2 (\zeta _{-,2} \zeta _{+,2}-\zeta _{-,3} \zeta _{+,3}-2)-4 \tilde{u} (\zeta _{-,1} \zeta _{+,1}+\zeta _{-,2} \zeta _{+,2}-\zeta _{-,3} \zeta _{+,3}-1)  \nonumber\\
		&+4 (3 \zeta _{-,1} \zeta _{+,1}+3 \zeta _{-,2} \zeta _{+,2}+\zeta _{-,3} \zeta _{+,3}+1)) \tilde{s}+2 (8 \zeta _{-,3} \zeta _{+,3} \tilde{t}^2+8 \tilde{t}^2-16 \tilde{u} \zeta _{-,1} \zeta _{+,1} \tilde{t}+16 \zeta _{-,1} \zeta _{+,1} \tilde{t}-16 \tilde{u} \zeta _{-,2} \zeta _{+,2} \tilde{t}+16 \zeta _{-,2} \zeta _{+,2} \tilde{t}  \nonumber\\
		&+\sqrt{(\tilde{s}-4) \tilde{s}} \sqrt{-(\tilde{s}-4) (4 m^2 (\tilde{s}-4)-\tilde{s}^2+4 \tilde{s}+(\tilde{t}-\tilde{u})^2)} \zeta _{-,3} \zeta _{+,2} \tilde{t}-(\tilde{s}-4)\sqrt{-\tilde{s} (4 m^2 (\tilde{s}-4)-\tilde{s}^2+4 \tilde{s}+(\tilde{t}-\tilde{u})^2)} \zeta _{-,2} \zeta _{+,3} \tilde{t}  \nonumber\\
		&-16 \zeta _{-,3} \zeta _{+,3} \tilde{t}-16 \tilde{t}+8 \tilde{u}^2-16 \tilde{u}+16 \tilde{u} \zeta _{-,1} \zeta _{+,1}-16 \zeta _{-,1} \zeta _{+,1}+16 \tilde{u} \zeta _{-,2} \zeta _{+,2}-16 \zeta _{-,2} \zeta _{+,2}-\sqrt{-\tilde{s}(4 m^2 (\tilde{s}-4)-\tilde{s}^2+4 \tilde{s}+(\tilde{t}-\tilde{u})^2)} \nonumber\\
		&\times(\tilde{s}-4) \tilde{u} \zeta _{-,3} \zeta _{+,2}+\sqrt{(\tilde{s}-4) \tilde{s}} \sqrt{-(\tilde{s}-4) (4 m^2 (\tilde{s}-4)-\tilde{s}^2+4 \tilde{s}+(\tilde{t}-\tilde{u})^2)} \tilde{u} \zeta _{-,2} \zeta _{+,3}+8 \tilde{u}^2 \zeta _{-,3} \zeta _{+,3}-16 \tilde{u} \zeta _{-,3} \zeta _{+,3}+16 \zeta _{-,3} \zeta _{+,3} \nonumber\\
		&+16)))  ~,\\
		H_{23}&=\frac{16m}{(\tilde{s}-4)^{2} (\tilde{s}-4 m^2) \sqrt{-(4 m^2 (\tilde{s}-4)-\tilde{s}^2+4 \tilde{s}+(\tilde{t}-\tilde{u})^2)}}\sqrt{\tilde{s}}(-\tilde{s}^2 (2 \sqrt{(\tilde{s}-4) \tilde{s}} \sqrt{-(\tilde{s}-4) (4 m^2 (\tilde{s}-4)-\tilde{s}^2+4 \tilde{s}+(\tilde{t}-\tilde{u})^2)}  \nonumber\\
		&\times(\zeta _{-,2} \zeta _{+,3}-\zeta _{+,2} \zeta _{-,3})+\tilde{t}^3 (\zeta _{-,1} \zeta _{+,1}-\zeta _{-,2} \zeta _{+,2}-\zeta _{-,3} \zeta _{+,3}+1)-3 \tilde{t}^2 \tilde{u} (\zeta _{-,1} \zeta _{+,1}-\zeta _{-,2} \zeta _{+,2}-\zeta _{-,3} \zeta _{+,3}+1)+3 \tilde{t} \tilde{u}^2 (\zeta _{-,1} \zeta _{+,1}-\zeta _{-,2} \zeta _{+,2} \nonumber\\
		&-\zeta _{-,3} \zeta _{+,3}+1)-16 \tilde{t} (3 \zeta _{-,1} \zeta _{+,1}+\zeta _{-,2} \zeta _{+,2}+\zeta _{-,3} \zeta _{+,3}+3)+\tilde{u}^3 (-\zeta _{-,1} \zeta _{+,1}+\zeta _{-,2} \zeta _{+,2}+\zeta _{-,3} \zeta _{+,3}-1)+16 \tilde{u} (3 \zeta _{-,1} \zeta _{+,1}+\zeta _{-,2} \zeta _{+,2} \nonumber\\
		&+\zeta _{-,3} \zeta _{+,3}+3))+8 \tilde{s}(\sqrt{(\tilde{s}-4) \tilde{s}} \sqrt{-(\tilde{s}-4) (4 m^2 (\tilde{s}-4)-\tilde{s}^2+4 \tilde{s}+(\tilde{t}-\tilde{u})^2)} (\zeta _{-,2} \zeta _{+,3}-\zeta _{+,2} \zeta _{-,3})-8 \tilde{t} (\zeta _{-,1} \zeta _{+,1}+\zeta _{-,2} \zeta _{+,2} \nonumber\\
		&+\zeta _{-,3} \zeta _{+,3}+1)+8 \tilde{u}(\zeta _{-,1} \zeta _{+,1}+\zeta _{-,2} \zeta _{+,2}+\zeta _{-,3} \zeta _{+,3}+1)+\tilde{t}^3 (\zeta _{-,1} \zeta _{+,1}+1)-3 \tilde{t}^2 (\zeta _{-,1} \zeta _{+,1} \tilde{u}+\tilde{u})+3 \tilde{t} \tilde{u}^2(\zeta _{-,1} \zeta _{+,1}+1)-\tilde{u}^3 (\zeta _{-,1} \zeta _{+,1} \nonumber\\
		&+1))-4 (\tilde{t}-\tilde{u})^2 (\sqrt{(\tilde{s}-4) \tilde{s}} \sqrt{-(\tilde{s}-4) (4 m^2 (\tilde{s}-4)-\tilde{s}^2+4 \tilde{s}+(\tilde{t}-\tilde{u})^2)}(\zeta _{+,2} \zeta _{-,3}-\zeta _{-,2} \zeta _{+,3})+4 \tilde{t}(\zeta _{-,1} \zeta _{+,1}+\zeta _{-,2} \zeta _{+,2}+\zeta _{-,3} \zeta _{+,3} \nonumber\\
		&+1)-4 \tilde{u}(\zeta _{-,1} \zeta _{+,1}+\zeta _{-,2} \zeta _{+,2}+\zeta _{-,3} \zeta _{+,3}+1))-4 m^2 (\tilde{s}-4) (2 (\tilde{s}-4) \sqrt{-\tilde{s}(4 m^2 (\tilde{s}-4)-\tilde{s}^2+4 \tilde{s}+(\tilde{t}-\tilde{u})^2)} (\zeta _{+,2} \zeta _{-,3}-\zeta _{-,2} \zeta _{+,3}) \nonumber\\
		&+(\tilde{s}-4) \tilde{t}(\tilde{s} (\zeta _{-,1} \zeta _{+,1}-\zeta _{-,2} \zeta _{+,2}-\zeta _{-,3} \zeta _{+,3}+1)-4 (\zeta _{-,1} \zeta _{+,1}+\zeta _{-,2} \zeta _{+,2}+\zeta _{-,3} \zeta _{+,3}+1))-(\tilde{s}-4 )\tilde{u}(\tilde{s} (\zeta _{-,1} \zeta _{+,1}-\zeta _{-,2} \zeta _{+,2}-\zeta _{-,3} \zeta _{+,3} \nonumber\\
		&+1)-4 (\zeta _{-,1} \zeta _{+,1}+\zeta _{-,2} \zeta _{+,2}+\zeta _{-,3} \zeta _{+,3}+1)))+\tilde{s}^4 (\tilde{t}-\tilde{u}) (\zeta _{-,1} \zeta _{+,1}-\zeta _{-,2} \zeta _{+,2}-\zeta _{-,3} \zeta _{+,3}+1)-4 \tilde{s}^3 (\tilde{t}-\tilde{u}) (3 \zeta _{-,1} \zeta _{+,1}-\zeta _{-,2} \zeta _{+,2}-\zeta _{-,3} \zeta _{+,3}+3))~,\\
		H_{31}&=(8 \tilde{s}(\tilde{s}^2 \sqrt{-(\tilde{s}-4) (m^4-2 m^2 (\tilde{t}+1)+(\tilde{s}-2) \tilde{t}+\tilde{t}^2+1)} (\zeta _{-,1} \zeta _{+,3}+\zeta _{+,1} \zeta _{-,3})-2 \tilde{s} (-\zeta _{-,1} \tilde{t} (\sqrt{-(m^4-2 m^2 (\tilde{t}+1)+(\tilde{s}-2) \tilde{t}+\tilde{t}^2+1)} \nonumber\\
		&\times\zeta _{+,3}\sqrt{(\tilde{s}-4)}+2 \sqrt{(\tilde{s}-4) \tilde{s}} \zeta _{+,2})+(m^2+1) \zeta _{-,1} \zeta _{+,3} \sqrt{-(m^4-2 m^2 (\tilde{t}+1)+(\tilde{s}-2) \tilde{t}+\tilde{t}^2+1)}\sqrt{(\tilde{s}-4)}+\zeta _{+,1} \zeta _{-,3} (m^2-\tilde{t}+1) \nonumber\\
		&\times\sqrt{-(\tilde{s}-4) (m^4-2 m^2 (\tilde{t}+1)+(\tilde{s}-2) \tilde{t}+\tilde{t}^2+1)})+4 \sqrt{(\tilde{s}-4) \tilde{s}} \zeta _{-,1} \zeta _{+,2} (m^4-2 m^2 (\tilde{t}+1)+(\tilde{t}-1)^2)-4 \sqrt{(\tilde{s}-4) \tilde{s}} \zeta _{+,1} \zeta _{-,2}\nonumber\\
		&\times(m^4-2 m^2 (\tilde{t}+1)+(\tilde{s}-2) \tilde{t}+\tilde{t}^2+1)))/((\tilde{s}-4) \sqrt{(4 m^2-\tilde{s})} \sqrt{m^4-2 m^2 (\tilde{t}+1)+(\tilde{s}-2) \tilde{t}+\tilde{t}^2+1})~,\\
		 H_{32}&=\frac{16m}{(\tilde{s}-4)^2(\tilde{s}-4 m^2)\sqrt{-(4 m^2 (\tilde{s}-4)-\tilde{s}^2+4\tilde{s}+(\tilde{t}-\tilde{u})^2)}}\sqrt{\tilde{s}}(\tilde{s}^2(2\sqrt{(\tilde{s}-4) \tilde{s}}\sqrt{-(\tilde{s}-4) (4 m^2 (\tilde{s}-4)-\tilde{s}^2+4 \tilde{s}+(\tilde{t}-\tilde{u})^2)} \nonumber\\
		&\times(\zeta _{-,2}\zeta _{+,3}-\zeta _{+,2}\zeta _{-,3})+\tilde{t}^3(\zeta _{-,1}\zeta _{+,1}-\zeta _{-,2}\zeta _{+,2}-\zeta _{-,3}\zeta _{+,3}+1)-3\tilde{t}^2\tilde{u}(\zeta _{-,1}\zeta _{+,1}-\zeta _{-,2}\zeta _{+,2}-\zeta _{-,3}\zeta _{+,3}+1)+3\tilde{t}\tilde{u}^2(\zeta _{-,1}\zeta _{+,1}-\zeta _{-,2}\zeta _{+,2} \nonumber\\
		&-\zeta _{-,3}\zeta _{+,3}+1)-16\tilde{t}(3\zeta _{-,1}\zeta _{+,1}+\zeta _{-,2} \zeta _{+,2}+\zeta _{-,3} \zeta _{+,3}+3)+\tilde{u}^3 (-\zeta _{-,1} \zeta _{+,1}+\zeta _{-,2} \zeta _{+,2}+\zeta _{-,3} \zeta _{+,3}-1)+16 \tilde{u} (3 \zeta _{-,1} \zeta _{+,1}+\zeta _{-,2} \zeta _{+,2}\nonumber\\
		&+\zeta _{-,3} \zeta _{+,3}+3))-8 \tilde{s}((\tilde{s}-4)\sqrt{-\tilde{s} (4 m^2 (\tilde{s}-4)-\tilde{s}^2+4 \tilde{s}+(\tilde{t}-\tilde{u})^2)} (\zeta _{-,2} \zeta _{+,3}-\zeta _{+,2} \zeta _{-,3})-8 \tilde{t} (\zeta _{-,1} \zeta _{+,1}+\zeta _{-,2} \zeta _{+,2}\nonumber\\
		&+\zeta _{-,3} \zeta _{+,3}+1)+8 \tilde{u} (\zeta _{-,1} \zeta _{+,1}+\zeta _{-,2} \zeta _{+,2}+\zeta _{-,3} \zeta _{+,3}+1)+\tilde{t}^3 (\zeta _{-,1} \zeta _{+,1}+1)-3 \tilde{t}^2 (\zeta _{-,1} \zeta _{+,1} \tilde{u}+\tilde{u})+3 \tilde{t} \tilde{u}^2 (\zeta _{-,1} \zeta _{+,1}+1)-\tilde{u}^3(\zeta _{-,1} \zeta _{+,1} \nonumber\\
		&+1))+4 (\tilde{t}-\tilde{u})^2 ((\tilde{s}-4)\sqrt{-\tilde{s}(4 m^2 (\tilde{s}-4)-\tilde{s}^2+4 \tilde{s}+(\tilde{t}-\tilde{u})^2)} (\zeta _{+,2} \zeta _{-,3}-\zeta _{-,2} \zeta _{+,3})+4 \tilde{t} (\zeta _{-,1} \zeta _{+,1}+\zeta _{-,2} \zeta _{+,2}+\zeta _{-,3} \zeta _{+,3}\nonumber\\
		&+1)-4 \tilde{u} (\zeta _{-,1} \zeta _{+,1}+\zeta _{-,2} \zeta _{+,2}+\zeta _{-,3} \zeta _{+,3}+1))+4 m^2 (\tilde{s}-4)(2(\tilde{s}-4) \sqrt{-\tilde{s}(4 m^2 (\tilde{s}-4)-\tilde{s}^2+4 \tilde{s}+(\tilde{t}-\tilde{u})^2)} (\zeta _{+,2} \zeta _{-,3}-\zeta _{-,2} \zeta _{+,3}) \nonumber\\
		&+(\tilde{s}-4) \tilde{t} (\tilde{s} (\zeta _{-,1} \zeta _{+,1}-\zeta _{-,2} \zeta _{+,2}-\zeta _{-,3} \zeta _{+,3}+1)-4 (\zeta _{-,1} \zeta _{+,1}+\zeta _{-,2} \zeta _{+,2}+\zeta _{-,3} \zeta _{+,3}+1))-(\tilde{s}-4 )\tilde{u} (\tilde{s} (\zeta _{-,1} \zeta _{+,1}-\zeta _{-,2} \zeta _{+,2}-\zeta _{-,3} \zeta _{+,3} \nonumber\\
		&+1)-4 (\zeta _{-,1} \zeta _{+,1}+\zeta _{-,2} \zeta _{+,2}+\zeta _{-,3} \zeta _{+,3}+1)))+\tilde{s}^4 (-\tilde{t}+\tilde{u}) (\zeta _{-,1} \zeta _{+,1}-\zeta _{-,2} \zeta _{+,2}-\zeta _{-,3} \zeta _{+,3}+1)+4 \tilde{s}^3( \tilde{t}-\tilde{u} )(3 \zeta _{-,1} \zeta _{+,1}-\zeta _{-,2} \zeta _{+,2} \nonumber\\
		&-\zeta _{-,3} \zeta _{+,3}+3))~,\\
		H_{33}&=\frac{1}{(\tilde{s}-4)^2 (-4 m^2+\tilde{s})} 8 (8 (\tilde{s}-4)^2 (\zeta _{-,1} \zeta _{+,1}+\zeta _{-,2} \zeta _{+,2}-\zeta _{-,3} \zeta _{+,3}-1) m^6-2 (\tilde{s}-4)^2 (\tilde{s} (5 \zeta _{-,1} \zeta _{+,1}-3 \zeta _{-,2} \zeta _{+,2}-5 \zeta _{-,3} \zeta _{+,3}+3)+4 (\tilde{t}  \nonumber\\
		&+\tilde{u}-2) (\zeta _{-,1} \zeta _{+,1}+\zeta _{-,2} \zeta _{+,2}-\zeta _{-,3} \zeta _{+,3}-1)) m^4+2 ((\zeta _{-,1} \zeta _{+,1}-\zeta _{-,2} \zeta _{+,2}-\zeta _{-,3} \zeta _{+,3}+1) \tilde{s}^4+(-6 \zeta _{-,1} \zeta _{+,1}+10 \zeta _{-,2} \zeta _{+,2}+14 \zeta _{-,3} \zeta _{+,3}\nonumber\\
		&+\tilde{t} (\zeta _{-,1} \zeta _{+,1}+\zeta _{-,2} \zeta _{+,2}-\zeta _{-,3} \zeta _{+,3}-1)+\tilde{u} (\zeta _{-,1} \zeta _{+,1}+\zeta _{-,2} \zeta _{+,2}-\zeta _{-,3} \zeta _{+,3}-1)-2) \tilde{s}^3+2 ((\zeta _{-,2} \zeta _{+,2}-1) \tilde{t}^2+2 ((\tilde{u}-3) \zeta _{-,1} \zeta _{+,1} \nonumber\\
		&-3 \zeta _{-,2} \zeta _{+,2}-\tilde{u} \zeta _{-,3} \zeta _{+,3}+3 \zeta _{-,3} \zeta _{+,3}+3) \tilde{t}+\tilde{u}^2 (\zeta _{-,2} \zeta _{+,2}-1)+2 (\zeta _{-,1} \zeta _{+,1}-7 \zeta _{-,2} \zeta _{+,2}-17 \zeta _{-,3} \zeta _{+,3}-9)-6 \tilde{u} (\zeta _{-,1} \zeta _{+,1}+\zeta _{-,2} \zeta _{+,2}\nonumber\\
		&-\zeta _{-,3} \zeta _{+,3}-1)) \tilde{s}^2-8 ((\zeta _{-,2} \zeta _{+,2}-\zeta _{-,3} \zeta _{+,3}-2) \tilde{t}^2+2 ((2 \tilde{u}-3) \zeta _{-,1} \zeta _{+,1}+(\tilde{u}-3) \zeta _{-,2} \zeta _{+,2}-\tilde{u} \zeta _{-,3} \zeta _{+,3}+3 \zeta _{-,3} \zeta _{+,3}+3) \tilde{t}+\tilde{u}^2 (\zeta _{-,2} \zeta _{+,2}  \nonumber\\
		&-\zeta _{-,3} \zeta _{+,3}-2)-6 \tilde{u} (\zeta _{-,1} \zeta _{+,1}+\zeta _{-,2} \zeta _{+,2}-\zeta _{-,3} \zeta _{+,3}-1)-16 (\zeta _{-,3} \zeta _{+,3}+1)) \tilde{s}-4(8 \zeta _{-,3} \zeta _{+,3} \tilde{t}^2+8 \tilde{t}^2-16 \tilde{u} \zeta _{-,1} \zeta _{+,1} \tilde{t}+16 \zeta _{-,1} \zeta _{+,1} \tilde{t} \nonumber\\
		&-16 \tilde{u} \zeta _{-,2} \zeta _{+,2} \tilde{t}+16 \zeta _{-,2} \zeta _{+,2} \tilde{t}+(\tilde{s}-4)\sqrt{-\tilde{s}(4 m^2 (\tilde{s}-4)-\tilde{s}^2+4 \tilde{s}+(\tilde{t}-\tilde{u})^2)} \zeta _{-,3} \zeta _{+,2} \tilde{t}-\sqrt{-\tilde{s}(4 m^2 (\tilde{s}-4)-\tilde{s}^2+4 \tilde{s}+(\tilde{t}-\tilde{u})^2)} \nonumber\\
		&\times(\tilde{s}-4) \zeta _{-,2} \zeta _{+,3} \tilde{t}-16 \zeta _{-,3} \zeta _{+,3} \tilde{t}-16 \tilde{t}+8 \tilde{u}^2-16 \tilde{u}+16 \tilde{u} \zeta _{-,1} \zeta _{+,1}-16 \zeta _{-,1} \zeta _{+,1}+16 \tilde{u} \zeta _{-,2} \zeta _{+,2}-16 \zeta _{-,2} \zeta _{+,2}-(\tilde{s}-4) \nonumber\\
		&\times\sqrt{-\tilde{s}(4 m^2 (\tilde{s}-4)-\tilde{s}^2+4 \tilde{s}+(\tilde{t}-\tilde{u})^2)} \tilde{u} \zeta _{-,3} \zeta _{+,2}+\sqrt{(\tilde{s}-4) \tilde{s}} \sqrt{-(\tilde{s}-4) (4 m^2 (\tilde{s}-4)-\tilde{s}^2+4 \tilde{s}+(\tilde{t}-\tilde{u})^2)} \tilde{u} \zeta _{-,2} \zeta _{+,3} \nonumber\\
		&+8 \tilde{u}^2 \zeta _{-,3} \zeta _{+,3}-16 \tilde{u} \zeta _{-,3} \zeta _{+,3}+16 \zeta _{-,3} \zeta _{+,3}+16)) m^2-\tilde{s} (2 (\zeta _{-,1} \zeta _{+,1}+\zeta _{-,2} \zeta _{+,2}+\zeta _{-,3} \zeta _{+,3}+1) \tilde{s}^3+((\zeta _{-,1} \zeta _{+,1}-\zeta _{-,3} \zeta _{+,3}) \tilde{t}^2 \nonumber\\
		&-2 (-\zeta _{-,2} \zeta _{+,2} \tilde{u}+\tilde{u}+\zeta _{-,1} \zeta _{+,1}+\zeta _{-,2} \zeta _{+,2}-\zeta _{-,3} \zeta _{+,3}-1) \tilde{t}+\tilde{u}^2 (\zeta _{-,1} \zeta _{+,1}-\zeta _{-,3} \zeta _{+,3})-2 \tilde{u} (\zeta _{-,1} \zeta _{+,1}+\zeta _{-,2} \zeta _{+,2}-\zeta _{-,3} \zeta _{+,3}-1) \nonumber\\
		&-2 (7 \zeta _{-,1} \zeta _{+,1}+7 \zeta _{-,2} \zeta _{+,2}+9 \zeta _{-,3} \zeta _{+,3}+9)) \tilde{s}^2-4 ((2 \zeta _{-,1} \zeta _{+,1}+\zeta _{-,2} \zeta _{+,2}-\zeta _{-,3} \zeta _{+,3}) \tilde{t}^2+2 (-2 \zeta _{-,1} \zeta _{+,1}-2 \zeta _{-,2} \zeta _{+,2}+2 \zeta _{-,3} \zeta _{+,3} \nonumber\\
		&+\tilde{u} (\zeta _{-,2} \zeta _{+,2}-\zeta _{-,3} \zeta _{+,3}-2)+2) \tilde{t}-4 \tilde{u} (\zeta _{-,1} \zeta _{+,1}+\zeta _{-,2} \zeta _{+,2}-\zeta _{-,3} \zeta _{+,3}-1)+\tilde{u}^2 (2 \zeta _{-,1} \zeta _{+,1}+\zeta _{-,2} \zeta _{+,2}-\zeta _{-,3} \zeta _{+,3})-4 (\zeta _{-,1} \zeta _{+,1} \nonumber\\
		&+\zeta _{-,2} \zeta _{+,2}+3 \zeta _{-,3} \zeta _{+,3}+3)) \tilde{s}+2 (8 \zeta _{-,1} \zeta _{+,1} \tilde{t}^2+8 \zeta _{-,2} \zeta _{+,2} \tilde{t}^2-16 \tilde{t} \tilde{u}-16 \zeta _{-,1} \zeta _{+,1} \tilde{t}-16 \zeta _{-,2} \zeta _{+,2} \tilde{t}+\sqrt{-\tilde{s}(4 m^2 (\tilde{s}-4)-\tilde{s}^2+4 \tilde{s}+(\tilde{t}-\tilde{u})^2)}  \nonumber\\
		&\times(\tilde{s}-4) \zeta _{-,3} \zeta _{+,2} \tilde{t}-\sqrt{(\tilde{s}-4) \tilde{s}} \sqrt{-(\tilde{s}-4) (4 m^2 (\tilde{s}-4)-\tilde{s}^2+4 \tilde{s}+(\tilde{t}-\tilde{u})^2)} \zeta _{-,2} \zeta _{+,3} \tilde{t}-16 \tilde{u} \zeta _{-,3} \zeta _{+,3} \tilde{t}+16 \zeta _{-,3} \zeta _{+,3} \tilde{t}+16 \tilde{t}+16 \tilde{u}\nonumber\\
		&+8 \tilde{u}^2 \zeta _{-,1} \zeta _{+,1}-16 \tilde{u} \zeta _{-,1} \zeta _{+,1}+16 \zeta _{-,1} \zeta _{+,1} +8 \tilde{u}^2 \zeta _{-,2} \zeta _{+,2}-16 \tilde{u} \zeta _{-,2} \zeta _{+,2}+16 \zeta _{-,2} \zeta _{+,2}-(\tilde{s}-4) \sqrt{-\tilde{s} (4 m^2 (\tilde{s}-4)-\tilde{s}^2+4 \tilde{s}+(\tilde{t}-\tilde{u})^2)} \nonumber\\
		&\times \tilde{u} \zeta _{-,3} \zeta _{+,2}+(\tilde{s}-4) \sqrt{-\tilde{s}(4 m^2 (\tilde{s}-4)-\tilde{s}^2+4 \tilde{s}+(\tilde{t}-\tilde{u})^2)} \tilde{u} \zeta _{-,2} \zeta _{+,3}+16 \tilde{u} \zeta _{-,3} \zeta _{+,3}-16 \zeta _{-,3} \zeta _{+,3}-16)))~.
	\end{align}
\end{subequations}
\end{widetext}

\bibliography{ref}

\end{document}